\documentclass[twocolumn,prl,floatfix]{revtex4-1}
\usepackage[section]{placeins}
\usepackage[utf8]{inputenc}
\usepackage{graphicx}
\usepackage{color}
\usepackage{amsmath}
\usepackage[left,running]{lineno}
\usepackage[utf8]{inputenc}
\usepackage{float}
\usepackage{enumitem}
\usepackage{multirow}
\newcommand{\bea}{\begin{eqnarray}} 
\newcommand{\eea}{\end{eqnarray}}
\newcommand{\beq}{\begin{equation}}
\newcommand{\eeq}{\end{equation}}
\newcommand{\pbp}{\langle\bar\psi\psi\rangle}
\newcommand \hmu {\hat{\mu}}
\newcommand\avr[1]{\left\langle{#1}\right\rangle}
\begin{document}
\title{The QCD crossover at finite chemical potential from lattice simulations}
\author{Szabolcs Borsanyi$^{a}$, Zoltan Fodor$^{a,b,c,d}$, Jana N. Guenther$^{a,e}$, Ruben Kara$^{a}$, Sandor D. Katz$^{b}$, Paolo Parotto$^{a}$, Attila Pasztor$^{b}$, Claudia Ratti$^{f}$, K\'alman K. Szab\'o$^{a,c}$}
\address{
$^a$ Department of Physics, Wuppertal University, Gaussstr.  20, D-42119\\
Wuppertal, Germany\\
$^b$ Inst.  for Theoretical Physics, ELTE E\"otv\"os Lor\' and University, P\'azm\'any P. s\'et\'any 1/A, H-1117 Budapest, Hungary\\
$^c$ J\"ulich Supercomputing Centre, Forschungszentrum J\"ulich, D-52425 J\"ulich, Germany\\
$^d$ Physics Department, UCSD, San Diego, CA 92093, USA\\
$^e$ University of Regensburg, Department of Physics, Regensburg D-93053, Germany \\
$^f$ Department of Physics, University of Houston, Houston, TX 77204, USA
}
\begin{abstract}
{
We provide the most accurate results for the QCD transition line so far.
We optimize the definition of the crossover temperature $T_c$, allowing for its
very precise determination, and extrapolate from imaginary chemical potential
up to real $\mu_B \approx 300$~MeV.
The definition of $T_c$ adopted in this work is based on the observation
that the chiral susceptibility as a function of the condensate is an almost
universal curve at zero and imaganiary $\mu_B$.
We obtain the parameters $\kappa_2=0.0153(18)$ and $\kappa_4=0.00032(67)$ as a
continuum extrapolation based on $N_t=10,12$ and $16$ lattices with physical
quark masses.
We also extrapolate the peak value of the chiral susceptibility
and the width of the chiral transition along the crossover line. 
In fact, both of these are consistent with a constant function of $\mu_B$.  We
see no sign of criticality in the explored range.
}
\end{abstract}
\maketitle

\emph{Introduction---} 
One of the most important open problems in the study of Quantum 
Chromodynamics (QCD) at finite temperature and density is the 
determination of the phase diagram of the theory in the 
temperature ($T$)-baryo-chemical potential ($\mu_B$) plane.
It is now established by first principle lattice QCD calculations that 
the transition at $\mu_B=0$ is a smooth crossover~\cite{Aoki:2006we, Bhattacharya:2014ara} for physical quark masses. 
Due to the lack of a real phase transition, the crossover temperature is of course 
ambiguous, since different definitions can lead to different values for it.
Observables related to chiral 
symmetry (i.e. the chiral condensate and its susceptibility) yield a 
transition temperature around $155 -160$~MeV
~\cite{Aoki:2006br,Aoki:2009sc,Borsanyi:2010bp,Bazavov:2011nk}.

Extending our knowledge to the $\mu_B>0$ part of the phase diagram
turns out to be very challenging, due to the notorious sign problem.
Since this makes direct simulation at finite $\mu_B$ impossible, 
the state-of-the-art for finite density QCD on fine lattices is to use one of two 
extrapolation methods. The first method is the direct calculation of Taylor coefficients
~\cite{Allton:2002zi,Allton:2005gk,Gavai:2008zr,Basak:2009uv,Borsanyi:2011sw,Borsanyi:2012cr,Bellwied:2015lba,Ding:2015fca,Bazavov:2017dus,Bazavov:2018mes,Bazavov:2020bjn} 
using simulations at $\mu_B=0$, while the second is to use
simulations at imaginary chemical potentials ($\mu_B^2<0$) where the sign problem is 
absent, and later perform an extrapolation of different quantities to a real chemical 
potential ($\mu_B^2>0$)~\cite{deForcrand:2002hgr,DElia:2002tig,DElia:2009pdy,Cea:2014xva,Bonati:2014kpa,Cea:2015cya,Bonati:2015bha,Bellwied:2015rza,DElia:2016jqh,Gunther:2016vcp,Alba:2017mqu,Vovchenko:2017xad,Bonati:2018nut,Borsanyi:2018grb}.
It is often conjectured that in the $(T,\mu_B)$ plane the  
crossover line, departing from $(T_c,\mu_B=0)$, eventually 
turns into a first-order transition line. 
The point $(T_{\rm CEP},\mu_{\rm CEP})$ separating the crossover and the 
first-order transitions is known as the critical endpoint (CEP), where the 
transition is expected to be of second order. 
Though there have been attempts in extracting information about the location
of the supposed CEP from lattice simulations~\cite{Fodor:2001pe,Fodor:2004nz,DElia:2016jqh,Bazavov:2017dus,Fodor:2019ogq,Giordano:2019slo,Mukherjee:2019eou,Giordano:2019gev}, these attempts face great 
difficulties, as extrapolation-type methods have the property that they give
reliable results mostly in the immediate vicinity of $\mu_B=0$. 

\begin{figure*}[ht]
\begin{center}
\hbox to \hsize{
\includegraphics[width=2.34in]{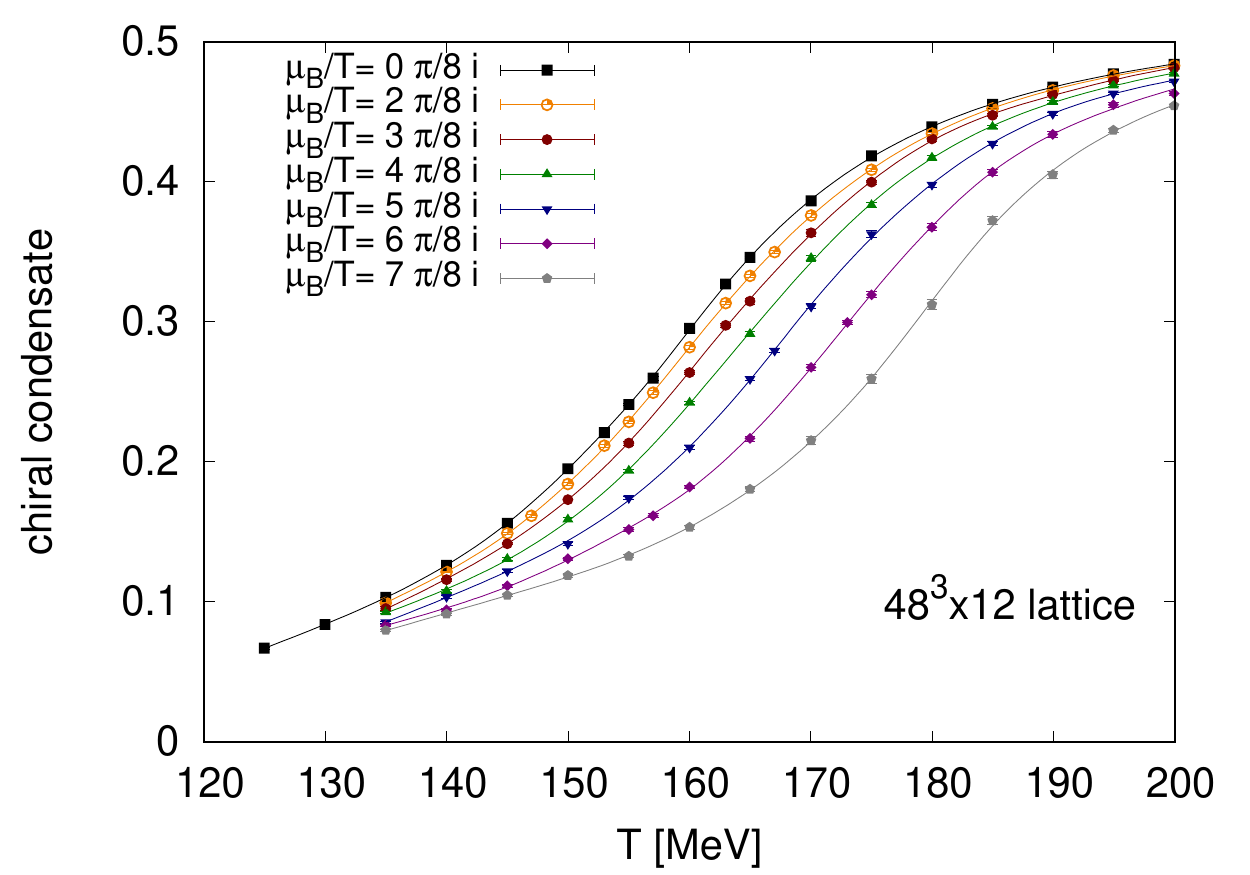}
\includegraphics[width=2.34in]{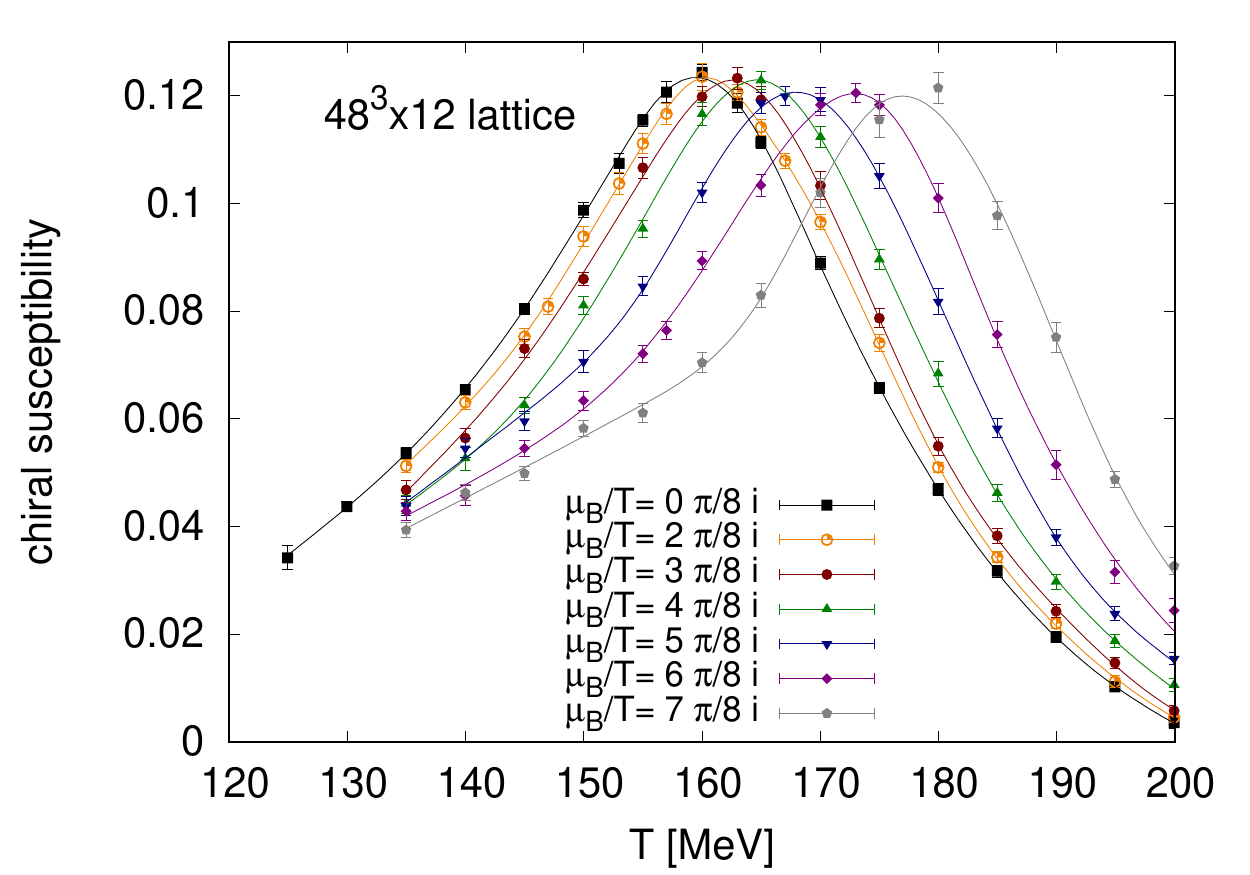}
\includegraphics[width=2.34in]{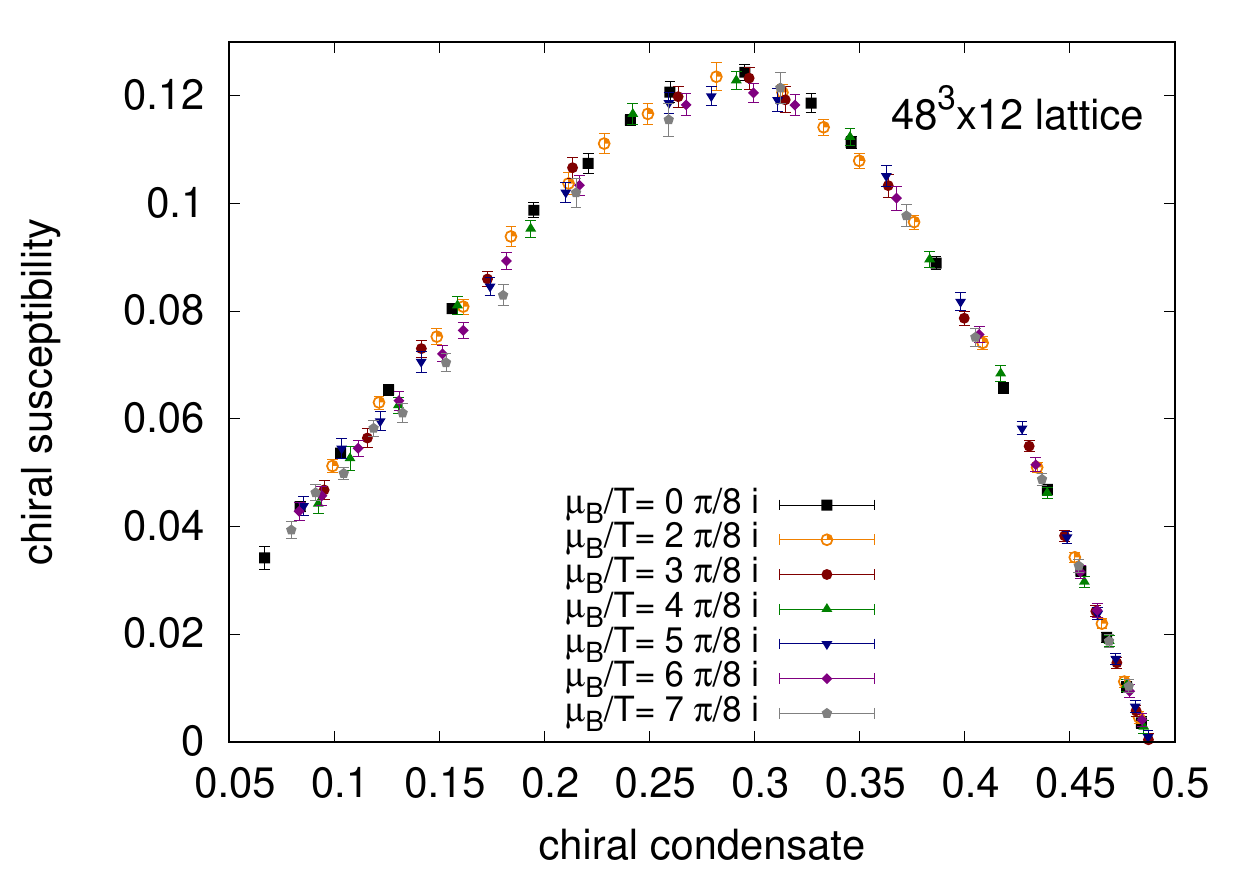}
\vspace{-0.8cm}
}
\end{center}
\caption{\label{fig:chinorm}
Renormalized chiral condensate $\pbp$ (left) and chiral susceptibility 
$\chi$ (middle) as functions of the temperature for the intermediate lattice 
spacing in this study. The black curves correspond to vanishing baryon density, 
while results for various imaginary values of the chemical
potential are shown in other colors. Finally, in the right panel we show the susceptibility as a function of the condensate. In this representation the chemical potential dependence
is very weak.
}
\end{figure*}

In this letter, we address the problem of calculating
the Taylor coefficients of the crossover temperature around $\mu_B=0$, parametrized as:
\begin{equation}
\frac{T_c(\mu_B)}{ T_c(\mu_B=0)}=1
- \kappa_2 \left(\frac{\mu_B}{T_c(\mu_B)}\right)^2
- \kappa_4 \left(\frac{\mu_B}{T_c(\mu_B)}\right)^4\dots
\label{eq:kappa}
\end{equation}
along the phenomenologically relevant strangeness neutrality line. In this work we improve the uncertainty on $\kappa_4$ available in
the literature~\cite{Bazavov:2018mes} by a factor of $6$, giving a state-of-the-art determination
of the cross-over line in the $(T,\mu_B)$ plane. 
In particular, as we will show, at 
chemical potentials $\mu_B>200~\mathrm{MeV}$ the error
on the $T_c$ extrapolation is dominated by the sub-leading coefficients
e.g. $\kappa_4$.
The coefficients $\kappa_2$ and $\kappa_4$ can be calculated with either one
of the standard extrapolation methods. A direct evaluation of
the $\mu_B$ derivatives from $\mu_B=0$ ensembles was used in Refs.~\cite{Kaczmarek:2011zz, Endrodi:2011gv}.
The current state-of-the art using the $\mu_B=0$ simulation method is 
Ref.~\cite{Bazavov:2018mes}, which includes the
first continuum extrapolated results for $\kappa_4$. 
Here we will employ an analytical continuation from imaginary $\mu_B$ instead, and 
use lattices as fine as  $N_t = 16$.
This is motivated by the fact that the signal/noise ratio of higher $\mu_B$ derivatives is
suppressed with powers of the lattice volume, therefore the calculation of higher 
order derivatives requires very high statistics. Determinations of $\kappa_2$ using 
the imaginary $\mu_B$ method with continuum
extrapolation include Refs.~\cite{Bonati:2015bha,Bellwied:2015rza}. Finally, in 
Ref.~\cite{Bonati:2018nut} the two methods were compared with a careful check of the systematics, 
and a very good agreement was found for the coefficient $\kappa_2$. 

We also study the strength of the crossover by extrapolating the width of the transition
and the value of the chiral susceptibility at the transition to real $\mu_B$ in the
continuum limit. While one always has to be careful not to over-interpret results from extrapolations, we currently do not see any sign of criticality up to $\mu_B \approx 300$~MeV, as the crossover transition does not get narrower or stronger in this region.

\vspace{0.1in}
\emph{On chiral observables in the transition region.---} 
For the lattice simulations we use 4-stout improved staggered fermions with an aspect ratio of $LT=4$ and temporal
          lattice sizes of $N_t=10,12,16$. The details of the simulation setup can be found in the supplemental material.
          
The main observables in this study are the renormalized dimensionless chiral condensate and susceptibility, respectively defined as:
\begin{equation}
    \begin{aligned}
        \pbp &= - \left[ \pbp_T - \pbp_0 \right]\frac{m_{\rm ud}}{f_\pi^4} \, ,   \\
        \chi &= \left[ \chi_T - \chi_0 \right]\frac{m_{\rm ud}^2}{f_\pi^4} \, \, , \quad \text{with} \\ 
        \pbp_{T,0}   &= \frac{T}{V} \frac{\partial \log Z}  {\partial m_{\rm ud}} \quad \chi_{T,0} = \frac{T}{V} \frac{\partial^2 \log Z}{\partial m_{\rm ud}^2}  \rm{,}
    \end{aligned}
\label{eq:chiobs}
\end{equation}
where we assumed isospin symmetry, i.e. $m_u=m_d=m_{ud}$. In the above equations, the subscripts $T,0$ indicate values at finite- and zero-temperature, respectively. In the following, $\pbp$ and $\chi$ are always shown after applying the correction to satisfy $n_s=0$ with zero statistical error (see the supplemental material for details). The peak height of the susceptibility is an indicator for the strength of the
transition, while the peak position in temperature serves as a definition for the
chiral cross-over temperature. It was pointed out in
Refs.~\cite{Aoki:2006br,Aoki:2009sc} that different normalizations of the
susceptibility, such as using $1/f_\pi^4$ or $1/T^4$ to define $\chi$ in
Eq.~(\ref{eq:chiobs}) can shift the peak position by 11~MeV. This difference
could be considered as a measure for the broadness of the chiral transition.

Our normalization choice in Eq.~(\ref{eq:chiobs}) was motivated 
by two observations, shown in Fig.~\ref{fig:chinorm} and explained below. These observations (together with
the improved statistics and the more accurate tuning of $\mu_S(\mu_B)$ to $n_S=0$)
allow a very precise determination of $T_c$ as a function of 
imaginary chemical potential, which in turn allows a precise 
determination of the parameters $\kappa_2$ and $\kappa_4$.
We explored the chiral condensate 
and susceptibility in a broad range of imaginary baryo-chemical potential. 
In all panels of Fig.~\ref{fig:chinorm}, the black curves correspond to $\mu_B=0$. 
In the left and middle panel we show the chiral condensate and susceptibility as 
functions of the temperature. By construction, our
renormalized condensate is zero at $T=0$ and positive at high temperature, 
because of the explicit vacuum subtraction and the overall negative sign in
Eq.~(\ref{eq:chiobs}). In both panels, one can observe the shifting of the
transition towards higher temperatures when an imaginary chemical potential
is introduced. In the right panel we show the susceptibility as a function
of the condensate. Here we converted the statistical error on the condensate into
an additional error on the susceptibility, by solving for 
$\pbp \left(T\right)={\rm const.}$ and substituting
the resulting $T$ into $\chi (T)$ (also taking the
correlation of the statistical errors into account).
Our first observation on the right panel of Fig.~\ref{fig:chinorm} is
that the form of the $\chi(\pbp)$ curve
is simpler than that of $\chi(T)$: a low (e.g. third or fourth)
order polynomial can fit the entire transition range with an excellent fit
quality. The second observation is that there is virtually no
chemical potential dependence in the $\chi(\pbp)$
function. This way the susceptibility can be modeled as a low order
polynomial of two variables, $\pbp$ and $\hmu=\mu_B/T$.
Had we used a different normalization for the susceptibility,
e.g.  $\chi(T) f_\pi^4/T^4$ as we did in Ref.~\cite{Borsanyi:2010bp},
the peak height would be strongly $\mu_B$-dependent and the collapse
of the $\chi(\pbp)$ curves at different (imaginary) chemical potentials would 
not happen. 

\begin{figure}[ht]
\begin{center}
\includegraphics[width=3.5in]{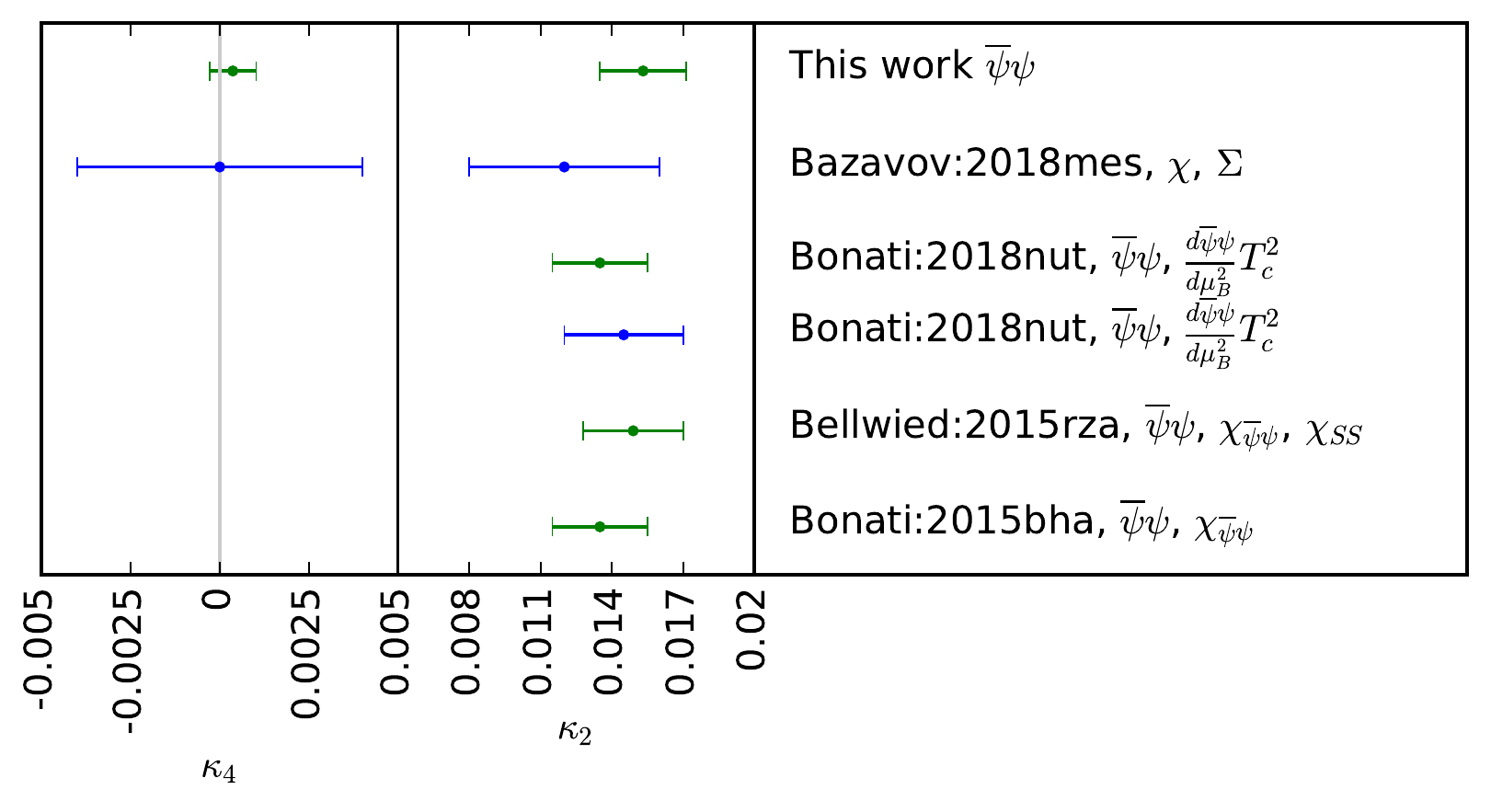}
\end{center}
\vspace{-0.8cm}
\caption{\label{fig:kappacmp}
Compilation of $\kappa_4$ (left) and $\kappa_2$ (right) coefficients from recent lattice
studies. We only include those papers where physical quark masses were used, a controlled continuum extrapolation was performed, and either strangeness
neutrality or $\mu_s=0$ was considered~\footnote{Note that while $\mu_s=0$ implies $\mu_S=\mu_B/3$ for all values of $\mu_B$, strangeness neutrality implies $\mu_S \approx \mu_B/4$ for small values of $\mu_B$.}.  The colors encode the numerical
approach. Blue points indicate simulations at $\mu_B=0$ only, where the $\mu_B$
dependence of $T_c$ was extracted using a Taylor expansion. The green points refer to works where imaginary chemical potentials were used.
}
\end{figure}

\begin{figure}[ht]
\begin{center}
\includegraphics[width=3.6in]{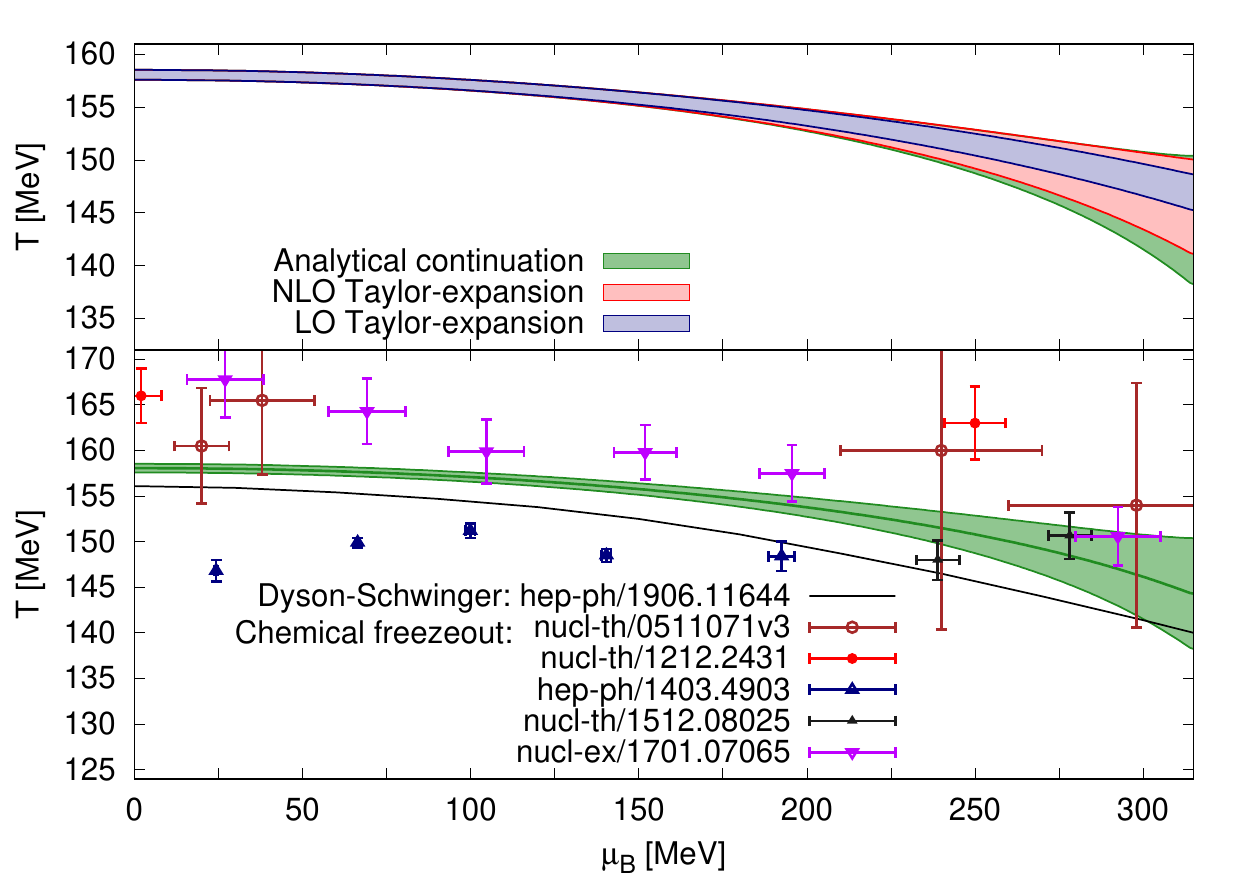}
\end{center}
\vspace{-0.8cm}
\caption{\label{fig:tcline}
Top: Transition line extrapolated from lattice simulations at imaginary
chemical potential using an analytical continuation with the ans\"atze used in
step iv) of our analysis (green band) compared with an extrapolation using the
formula in Eq.~(\ref{eq:kappa}) up to the order of $\kappa_4$ (red band) or up
to $\kappa_2$ (blue band).  The proximity of the full and NLO result suggests
that the higher order corrections are small in the range of $\mu_B$
considered here. 
Note that considering only the error bar of $\kappa_2$ underestimates the
full error.
The numerical values for the final analytical continuation, together with its
error, are tabulated in the supplemental material.
Bottom: Crossover line from the lattice compared with a prediction from
truncated Dyson-Schwinger equations ~\cite{Isserstedt:2019pgx} and some
estimates of the chemical freezeout parameters in heavy ion
collisions~\cite{Andronic:2005yp,Becattini:2012xb,Alba:2014eba,Vovchenko:2015idt,Adamczyk:2017iwn}.
Note that the width of the green band is not a representation of the width of
the crossover region, it depicts the statistical and systematic errors
achievable with the particular definition of the crossover temperature $T_c$
adopted in this work. Note also that the definition of the crossover
temperature adopted in Ref.~\cite{Isserstedt:2019pgx} is different from the one
used in this work.
}
\end{figure}

\begin{figure*}[ht]
\begin{center}
\includegraphics[width=3.3in]{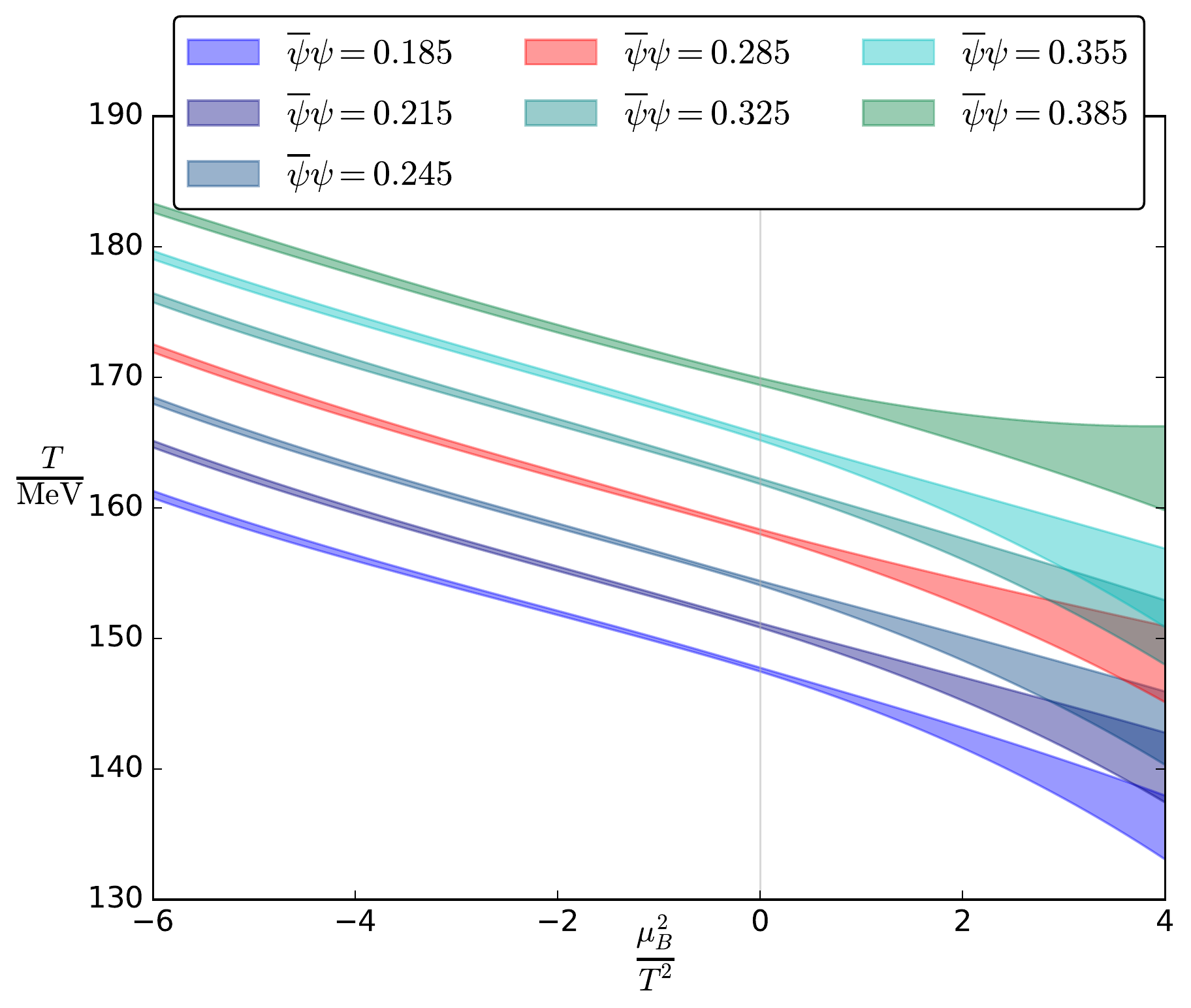} \hfill
\includegraphics[width=3.3in]{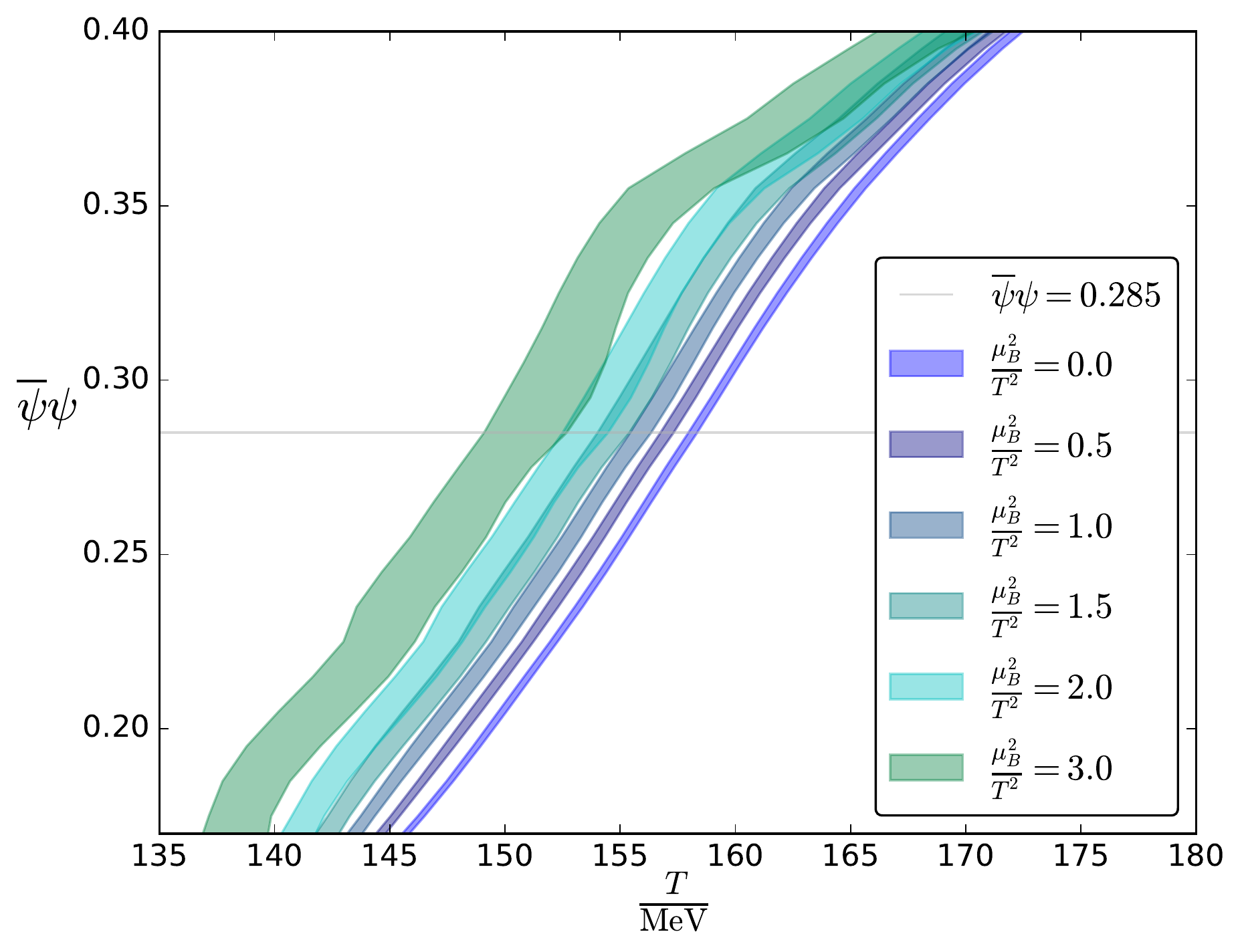} \vspace{-0.3in}
\end{center}
\caption{\label{fig:pbpextra}
Extrapolation of the width of the transition using contours of fixed values of
the renormalized chiral condensate. In the left panel we show the contours
for a set of fixed $\pbp$ values for both imaginary $\hmu_B$ (our simulation
domain) and for real $\hmu_B$, where the width of the band refers to
the combined extrapolation error. The red band roughly corresponds to
$T_c(\hmu_B^2)$. In the right panel we plot the same extrapolation in a
different representation: for fixed $\hmu_B$ we extrapolate the $\pbp$ values.
Here we get extrapolation errors on $T$, represented by the colored bands.
}
\end{figure*}

\vspace{0.1in}
\emph{The transition line and its analytical continuation.---}
Keeping the previous observations in mind, one can perform a precise
determination of $T_c$, as defined by the peak of $\chi$ in
Eq.~\eqref{eq:chiobs} for various values of the imaginary chemical potential.
$T_c(\mu_B^2)$ can then be fitted for the coefficients $\kappa_2$ and
$\kappa_4$.  This requires the following steps:
\setlist{nolistsep}
\begin{enumerate}[noitemsep, label=\roman*)]
    \item Determine the renormalized condensate $\pbp$ and susceptibility
$\chi$ in a two-dimensional parameter scan in $T$ and ${\rm Im} \mu_B$ using lattice simulations. 
Use these to obtain the susceptibility as a function of the condensate. 
    \item Search for the peak of $\chi(\pbp)$ through a low-order polynomial fit for each $N_t$ and ${\rm Im} \mu_B$ obtaining $\pbp_c(N_t, {\rm Im \mu_B})$.
    \item Use an interpolation of $\pbp(T)$ to convert the $\pbp_c$ to $T_c$
for each ${\rm Im \mu_B}/T$.
    \item Perform a global fit of $T_c(N_t, {\rm Im} \mu_B/T_c)$ to determine the coefficients 
    $\kappa_2$ and $\kappa_4$ for $1/N_t^2=0$. For this step we use various functions -- 
    all containing an independent $\kappa_6$ -- with coefficients depending linearly on 
    $1/N_t^2$. The choice of the fit functions is motivated by the mock data analysis presented in the supplemental material.
\end{enumerate}

The total systematic error comes from a pool of $256$ analyses: in step i) we
have two choices for the scale setting, two choices for the renormalization of
$\pbp$ and two for the renormalization of $\chi$; in step ii) we use two choices
for the fit function used to obtain the maximum of $\chi(\pbp)$ and two choices
for the fit range; in step iii) we use two different
interpolations of $\pbp(T)$; in step iv) we use two global fit functions and two
choices for the range in ${\rm Im }\mu_B/T$. This leads to a total of $2^8=256$
ways to analyze our lattice data. These results are combined with a uniform
weight.  More details on the analyses, the fit qualities and the error
estimates can be found in the supplemental material.
We finally obtain:
\begin{equation}
    \begin{aligned}
\kappa_2&=&0.0153  &\pm  0.0018  \, \, ,\\
\kappa_4&=&0.00032 &\pm 0.00067  \, \, .
    \end{aligned}
\end{equation}
We stress that the uncertainties on these two quantities are correlated.  We
put these results in the context of previous lattice studies in
Fig.~\ref{fig:kappacmp}.  The extrapolated value of $T_c(\mu_B)$ is shown in
Fig.~\ref{fig:tcline} (green band).
Note that the errors on $\kappa_2$ and $\kappa_4$ are dominated by the
statistical errors, as shown in the detailed discussion of the systematic error
estimate in the supplemental material. 

Since Ref.~\cite{Bellwied:2015rza} we have more than doubled the statistics
and introduced a more precise analysis. The overall error on $\kappa_2$ has
reduced slightly. The main result is the extraction of $\kappa_4$.
It appears to be a generic feature of deducing Taylor-coefficients
from polynomial fits: the increased precision on the input data leads to a
sensitivity to a higher order coefficient first, and only later to a reduction
of the error of both coefficients. This feature is also clearly seen in the
mock data analysis in the supplemental material. 

In Fig.~\ref{fig:tcline} we also show the comparison to the leading order
Taylor expansion result (using only $\kappa_2$) and the next to leading
order result (using $\kappa_2$ and $\kappa_4$). The latter is very close
to our full result (for $\mu_B<300~\mathrm{MeV}$), while the leading order
result has a much smaller uncertainty. Clearly, $\kappa_2$ is precise enough.
At intermediate $\mu_B$ the bottleneck for the precision of $T_c(\mu_B)$ 
is the error on $\kappa_4$. We also fitted $\kappa_6$, this turned out to
be small enough to be irrelevant for $\mu_B<300~\mathrm{MeV}$.

\vspace{0.1in}
\emph{Extrapolation of the transition width and strength.---} 
A natural definition of the width of the susceptibility peak is given by its 
second derivative at $T_c$ as $(\Delta T)^2 = -\chi(T_c) \left[ \frac{d^2}{dT^2} \chi \right]^{-1}_{T=T_c}$.
Unfortunately, evaluating this quantity is numerically difficult, so we introduce a simple width parameter $\sigma$ as a proxy for $\Delta T$ via:
\begin{equation}
    \begin{aligned}
\pbp(T_c \pm \sigma/2)&= \pbp_c \pm \Delta\pbp/2 \, \, ,
    \end{aligned}
\label{eq:Wdef}
\end{equation}
with $\pbp_c=0.285$ and $\Delta \pbp=0.14$. 
The choice of the range in $\pbp$ is such that it is consistent with a
linear behavior within our errorbars, meaning that the ratio $\Delta \pbp / \sigma$ 
can be used as a proxy for $\frac{d}{dT} \pbp |_{T=T_c}$ as well. The exact range in 
$\pbp$ is chosen such that $\sigma$ coincides with $\Delta T$ at zero and imaginary 
$\mu_B$. A more detailed discussion of the width parameter can be found in the
supplemental material.

\begin{figure}

\includegraphics[width=3.5in]{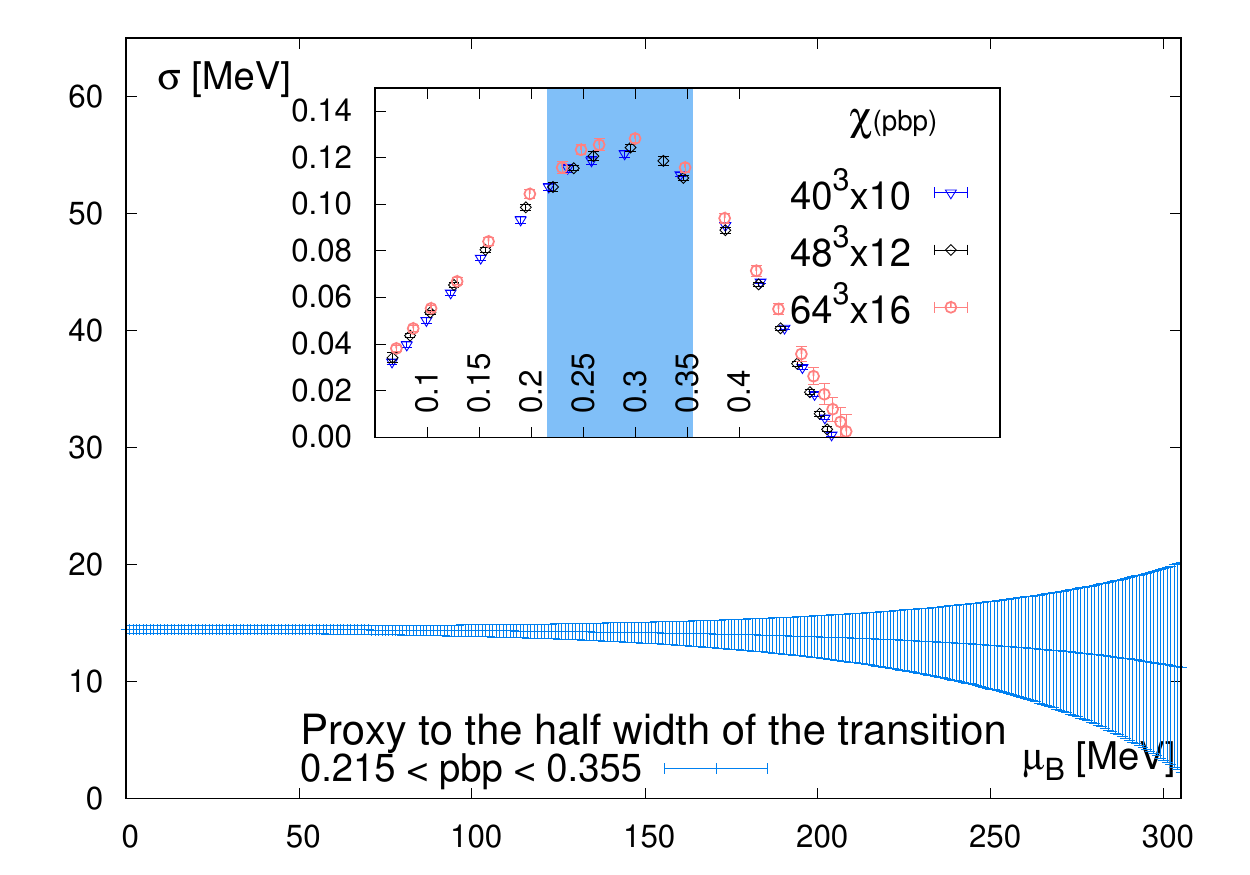}
 \vspace{0.1in} 
 
\includegraphics[width=3.5in]{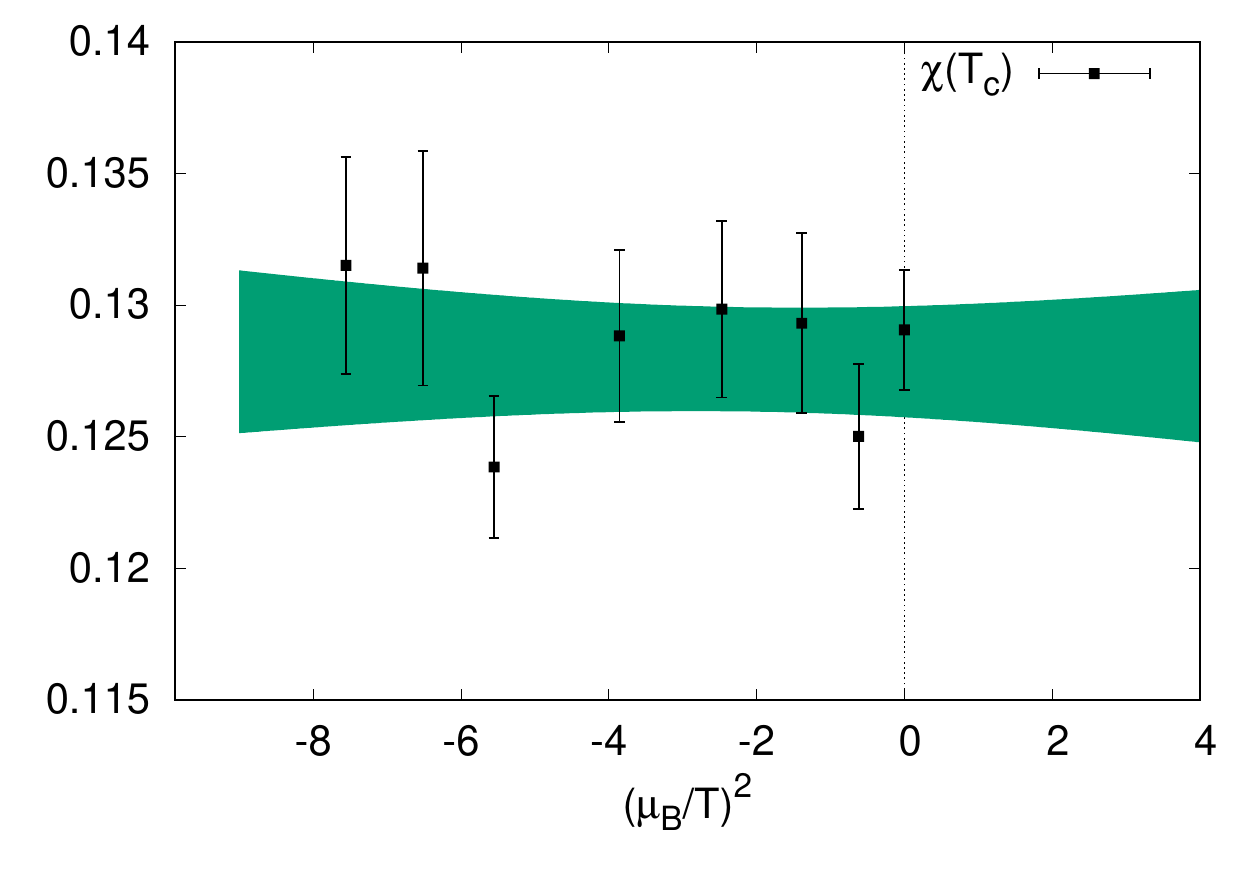}
\caption{\label{fig:chi_immu} 
Top: Half width $\sigma$ of the transition defined in Eq.~(\ref{eq:Wdef}) using the temperature difference of the contours $\pbp=0.31$ and $\pbp=0.19$. In the insert we show a plot of
the $\chi(\pbp)$ peak, where the shaded region corresponds to 
$\pbp_c \pm \Delta \pbp/2$. Both are extrapolated to real $\mu_B$. Bottom: Result 
of a $\mu_B$-by-$\mu_B$ analysis for the value of the chiral susceptibility at 
the crossover temperature after continuum extrapolation
and including the systematic errors for $LT_c=4$. The green band shows a 
linear extrapolation in $\hmu_B^2$.
}
\end{figure}

We can define, for any value of $\pbp$, the temperature function 
\begin{equation}
T^{\rm contour}_{\pbp=x} (\hmu_B) = T,~\mathrm{where}~\pbp(T,\hmu_B)=x \, \, .
\end{equation}
In the left panel of Fig.~\ref{fig:pbpextra} we show these contours in the 
$(T,\mu_B^2)$ plane for a selection
of $\pbp$ values. We show continuum extrapolations
and include the systematic errors. For this analysis the chiral
susceptibility plays no role, since we use the same interpolations of 
$\pbp(T)$ as in step iii)
of the analysis of the transition line, and two-dimensional fits (continuum
and in $\hmu_B^2$) analogous to those in step iv).
We conclude that the half width of the transition -- shown in the upper panel of Fig.~\ref{fig:chi_immu} -- is consistent with a constant up to
$\mu_B\approx 300~\mathrm{MeV}$ within the uncertainty from the extrapolation (we note that $50\%$ uncertainty is reached at $\mu_B \approx 280$~MeV). 
We also show $\pbp$ as a function of $T$ for several fixed values of 
real $\mu_B/T$ in the right panel of Fig.~\ref{fig:pbpextra}, as extrapolated using the
contours in the left panel of the same Figure.

Finally, as a proxy for the strength of the crossover, we study the value of the chiral susceptibility at the crossover temperature.
We get this for each ${\rm Im }\mu_B$ and $N_t$ as a byproduct of steps i-ii) of the analysis for $\kappa_2$ and $\kappa_4$. If one then
performs a continuum extrapolation of the resulting values for fixed values of ${\rm Im }\mu_B$, one gets the lower panel of Fig.~\ref{fig:chi_immu}. Again,
we see a very mild $\hmu_B^2$ dependence, consistent with a constant.

\emph{Summary and discussion.---} 
The main result of this work is a precise determination of the 
parameters $\kappa_2$ and $\kappa_4$ of the crossover line in
finite density QCD. For the determination of the crossover line,
we used the experimentally relevant $\mu_S(\mu_B)$
tuned to keep $n_S=0$. Based on the observation that the chiral susceptibility as a function
of the condensate is a rather simple function, only weakly dependent
on the imaginary chemical potential, we were able to obtain the transition
temperature as a function of the imaginary chemical potential to very high accuracy.
These pure lattice results can be used for further model building, and are summarized
in the supplemental material. The high precision data at imaginary $\mu_B$ in 
turn allowed us to fit the $\mu_B^2$ and
$\mu_B^4$ Taylor coefficients of the crossover temperature, $\kappa_2$ 
and $\kappa_4$.
In particular, while our determination of $\kappa_4$ is still consistent with zero,
the error is 6 times smaller than the one previously available in the literature, 
and therefore represents the state-of-the-art in the study of the phase diagram in the $(T,\mu_B)$ plane with current lattice techniques.
As a byproduct, we also obtain the most precise value for the central
temperature of the crossover at $\mu_B=0$ so far, as well as the width of the transition:
\begin{equation}
\begin{aligned}
T_c(LT=4, \mu_B=0)      &= &158.0 & & \pm & & 0.6~\mathrm{MeV} \\
\Delta T(LT=4, \mu_B=0) &= &15 & & \pm & & 1~\mathrm{MeV}
\end{aligned}
\end{equation}
The present definition was actually included in our earlier list of observables with $T_c(LT=3, \mu_B=0)=157(3)(3)~\mathrm{MeV}$ \cite{Aoki:2009sc}.  
Recently the HotQCD collaboration has published $T_c(LT=4, \mu_B=0)=156.5\pm 1.5~\mathrm{MeV}$ for the peak of the chiral susceptibility in Ref.~\cite{Bazavov:2018mes}.

Note that all results in this letter were obtained at a fixed aspect ratio of
$LT=4$. Though this result is consistenct with earlier works, also with $LT=3$,
finite volume corrections can potentially be relevant with the achieved
precision. We also did not take into account isospin breaking and QED effects.

We also studied the strength of the phase transition as a function of $\mu_B$ 
by extrapolating our proxy for the transition width and the peak of the chiral
susceptibility from imaginary chemical
potentials. Even though one has to be careful with extrapolations, we see no sign of the transition getting
stronger up to $\mu_B \approx 300$~MeV. 

\vspace{0.1in}
\emph{Acknowledgements.---} 
This project was funded by the DFG grant SFB/TR55. 
The project also received support from the BMBF Grant No. 05P18PXFCA.
This work was also supported by
the Hungarian National Research,  Development and Innovation Office, NKFIH
grants KKP126769 and K113034. A.P. is supported by the J. Bolyai Research
Scholarship of the Hungarian Academy of Sciences and by the \'UNKP-19-4 New
National Excellence Program of the Ministry for Innovation and Technology.
This  material is based upon  work  supported  by  the National  Science
Foundation under grants no. PHY-1654219 and by the U.S. DoE, 
Office  of  Science,  Office  of  Nuclear  Physics, within the framework of the
Beam Energy Scan Topical (BEST) Collaboration.  The authors gratefully
acknowledge the Gauss Centre for Supercomputing e.V.  (www.gauss-centre.eu) for
funding this project by providing computing time on the GCS Supercomputer
JURECA/Booster at J\"ulich Supercomputing Centre (JSC), on HAZELHEN at HLRS,
Stuttgart as well as on SUPERMUC-NG at LRZ, Munich.  We acknowledge PRACE for
awarding us access to Piz Daint hosted at CSCS, Switzerland.
C.R. also acknowledges the support from the Center of Advanced
Computing and Data Systems at the University of Houston. 


\newpage

\section*{Supplemental Material}

\subsection{\label{sec:lat} Details of the lattice setup} 

Throughout this work we use tree-level Symanzik improvement for the gauge
action and four levels of stout smearing in the staggered fermion action.  
We use three lattice spacings in this work, which are given in terms
of the (Euclidean) temporal resolution of the isotropic lattices: $a=1/(N_t
T)$, with $N_t=10,12$ and $16$.

The parameters of our discretization are tuned in such a way that
the measured pion and kaon masses are equal to
$135$~MeV and $495$~MeV, respectively, if we use the pion
decay constant $f_\pi=130.41~\mathrm{MeV}$
for scale setting \cite{Tanabashi:2018oca}. As an alternative scale
setting we use the Wilson flow scale $w_0$ \cite{Borsanyi:2012zs}. 
Continuum extrapolated results do not depend on the choice of 
the scale setting procedure, while results at finite lattice spacings do. For
example, near the transition on our $40^3\times10$ lattice, $w_0 f_\pi$ differs
by $2.5\%$ from the continuum value, thus all hadrons appear $2.5\%$ lighter
with the $w_0$ scale setting. On our finer lattices, $48^3 \times 12$ and $64^3
\times 16$ this difference reduces to $2\%$ and $1\%$, respectively, and
vanishes in the continuum extrapolated results. 
Should there be any small deviation
between the two $f_\pi$ and $w_0$ based continuum extrapolations, we consider
the difference as part of the systematic error.
The actual simulation
parameters as well as the bare parameters are given in
Ref.~\cite{Bellwied:2015lba}. 
This action has already been used to calculate 
the equation of state a $\mu_B=0$ \cite{Borsanyi:2016ksw},
fluctuations of conserved
charges \cite{Bellwied:2015lba} as well as the cross-correlators
\cite{Bellwied:2019pxh} and fugacity expansion
coefficients~\cite{Alba:2017mqu,Vovchenko:2017xad}.

For the purpose of renormalization in Eq.~\eqref{eq:chiobs} in the main text, 
we calculate the vacuum condensate and susceptibility, on large lattices with 
$Lm_\pi\approx 4$ and $N_t/N_x \gtrsim 1.3$, for 9 values of the 
gauge coupling, ranging over
$\beta=3.55 - 4.0126$, corresponding to $a=0.19 - 0.063~\mathrm{fm}$. 
In this range the bare condensate varies by an order of magnitude.
We interpolate between the simulated gauge couplings by fitting
the natural logarithm of the $T=0$ condensate with a polynomial of order four
or five -- with $\chi^2/n_{\rm dof} =3.52/4$ and $2.6/2$ respectively.
The bare susceptibility at $T=0$, as well as its logarithm, were fitted with a
second order polynomial -- with $\chi^2/n_{\rm ndof}=8.4/6$ and $9.3/6$ 
respectively. The two interpolations of the condensate and the two
interpolations of the susceptibility were varied independently in the
systematic error analysis.

In this work we generate lattice ensembles for fixed ratios of zero and 
imaginary $\mu_B/T$. With the notation $\hmu_B=\mu_B/T$ we select eight
values where we perform a temperature scan with our three lattice spacings.
\begin{equation}
\hmu_B^{(j)} = i j \pi/8 \,,\qquad j=0,2,3,4,5,6,6.5,7 \, .
\label{eq:immu}
\end{equation}

\begin{table}[!h]
\begin{tabular}{|c|c|c|c|c|c|c|c|c|}
\hline
\multicolumn{9}{|c|}{$40^3\times 10$ lattice}\\
\hline
\multirow{2}{*}{$T$~[MeV]} & \multicolumn{8}{|c|}{${\mu}_B^I/T$}\\
& 0.000 &0.785 & 1.178 & 1.570 & 1.963 & 2.356 & 2.553 & 2.749 \\
\hline
135 & 20159 & 2042 & 2518 & 3255 & 2384 & 2690 & 4373 & 3728 \\
140 & 15898 & 8904 & 2555 & 3260 & 2407 & 2692 & 4381 & 3815 \\
145 & 9638 & 10061 & 2609 & 3259 & 2425 & 4444 & 4545 & 3883 \\
150 & 9382 & 9710 & 7192 & 6951 & 4840 & 2729 & 4516 & 3839 \\
155 & 9663 & 6235 & 4812 & 9966 & 8654 & 2735 & 4382 & 5713 \\
160 & 9783 & 6223 & 4680 & 10128 & 9001 & 7695 & 4595 & 5577 \\
165 & 19507 & 11576 & 2799 & 9806 & 9774 & 10379 & 4676 & 5920 \\
170 & 16196 & 12332 & 5634 & 4226 & 10300 & 11591 & 4815 & 6035 \\
175 & 10593 & 13316 & 1540 & 7110 & 5287 & 11453 & 4875 & 4271 \\
180 & 10007 & 12950 & 1653 & 8313 & 2096 & 3279 & 5256 & 4501 \\
185 & 5492 & 1766 & 5959 & 6841 & 2235 & 3521 & 5666 & 4877 \\
190 & 9938 & 1855 & 1878 & 6891 & 2357 & 7636 & 6131 & 5240 \\
195 & 6951 & 1473 & 1155 & 3426 & 6087 & 7074 & 4823 & 3062 \\
200 & 9765 & 1518 & 2016 & 8160 & 6157 & 6609 & 5081 & 3244 \\
\hline
\hline
\multicolumn{9}{|c|}{$48^3\times 12$ lattice}\\
\hline
\multirow{2}{*}{$T$~[MeV]} & \multicolumn{8}{|c|}{${\mu}_B^I/T$}\\
& 0.000 &0.785 & 1.178 & 1.570 & 1.963 & 2.356 & 2.553 & 2.749 \\
\hline
135 & 27681 & 5925 & 2632 & 4247 & 3459 & 4067 & 5130 & 5312 \\
140 & 27723 & 5806 & 4051 & 4187 & 3471 & 4015 & 5174 & 5275 \\
145 & 27147 & 5677 & 9596 & 6914 & 6018 & 5125 & 5326 & 5397 \\
150 & 18137 & 5704 & 15529 & 7598 & 3587 & 6564 & 5445 & 5429 \\
155 & 27359 & 5939 & 7350 & 7651 & 8432 & 6540 & 5390 & 5670 \\
160 & 17460 & 6350 & 6888 & 7912 & 11561 & 9062 & 5386 & 5695 \\
165 & 27257 & 7043 & 5827 & 9574 & 13957 & 7982 & 5436 & 5826 \\
170 & 8833 & 7916 & 5527 & 6533 & 9055 & 9418 & 5621 & 6052 \\
175 & 16805 & 8777 & 4338 & 3912 & 5240 & 7888 & 5771 & 5965 \\
180 & 17182 & 9743 & 3367 & 4347 & 5924 & 7281 & 6230 & 6311 \\
185 & 14146 & 10649 & 3618 & 4583 & 6392 & 7750 & 6640 & 6805 \\
190 & 18668 & 11405 & 3851 & 4934 & 6847 & 3598 & 6982 & 7171 \\
195 & 14972 & 12223 & 4023 & 4730 & 6025 & 1702 & 6152 & 7541 \\
200 & 20991 & 12942 & 4258 & 5038 & 6325 & 1736 & 6608 & 7940 \\
\hline
\hline
\multicolumn{9}{|c|}{$64^3\times 16$ lattice}\\
\hline
\multirow{2}{*}{$T$~[MeV]} & \multicolumn{8}{|c|}{${\mu}_B^I/T$}\\
& 0.000 &0.785 & 1.178 & 1.570 & 1.963 & 2.356 & 2.553 & 2.749 \\
\hline
135 & 23194 & 1909 & 5659 & 6288 & 3927 & 4400 & 3100 & 3261 \\
140 & 13587 & 2813 & 6632 & 4915 & 4856 & 4352 & 3089 & 3238 \\
145 & 13682 & 2679 & 7157 & 5791 & 4713 & 5965 & 2991 & 3310 \\
150 & 13697 & 2577 & 9095 & 5777 & 4346 & 4286 & 2902 & 3406 \\
155 & 14164 & 2865 & 7886 & 6900 & 4706 & 4411 & 3005 & 3114 \\
160 & 14465 & 2689 & 9136 & 6870 & 4980 & 6124 & 3439 & 3129 \\
165 & 14983 & 7714 & 9809 & 7786 & 6572 & 7286 & 3673 & 3375 \\
170 & 15594 & 8360 & 12324 & 6378 & 5313 & 7205 & 3927 & 3256 \\
175 & 16362 & 9380 & 15056 & 6948 & 4911 & 8441 & 3382 & 3361 \\
180 & 16960 & 10453 & 8064 & 6966 & 5251 & 9173 & 3290 & 3546 \\
185 & 7689 & 3504 & 7844 & 7120 & 5723 & 8831 & 3602 & 3320 \\
190 & 33373 & 3416 & 5777 & 7543 & 6077 & 6306 & 4879 & 3678 \\
195 & 8918 & 4389 & 5931 & 7895 & 5841 & 6858 & 5204 & 3835 \\
200 & 14308 & 4770 & 6049 & 8336 & 5785 & 7289 & 2602 & 4035 \\
\hline
\end{tabular}

\caption{\label{tab:stat}
The number of analyzed configurations in the finite temperature ensembles.
}
\end{table}

\FloatBarrier


Several of these chemical potentials were already used in our earlier
work on the transition line \cite{Bellwied:2015rza}, where a key point was the
use of strangeness neutrality. This means that for each (imaginary)
baryo-chemical potential and temperature, the strangeness chemical potential
was tuned such that the expectation value of the strangeness density vanishes.
In this work, too, a non-zero value of $\mu_B$ always implies a matching
$\mu_S$ parameter with $n_S=0$. Thus, our extrapolations in $\mu_B$ also
extrapolate in $\mu_S$.

In Table~\ref{tab:stat} we list the number of analyzed configurations in the 
finite-temperature ensembles for the different lattices used in this work. 


\subsection{\label{sec:mock}Mock data analysis}

\begin{figure*} 
\begin{center}
\includegraphics[width=3.5in]{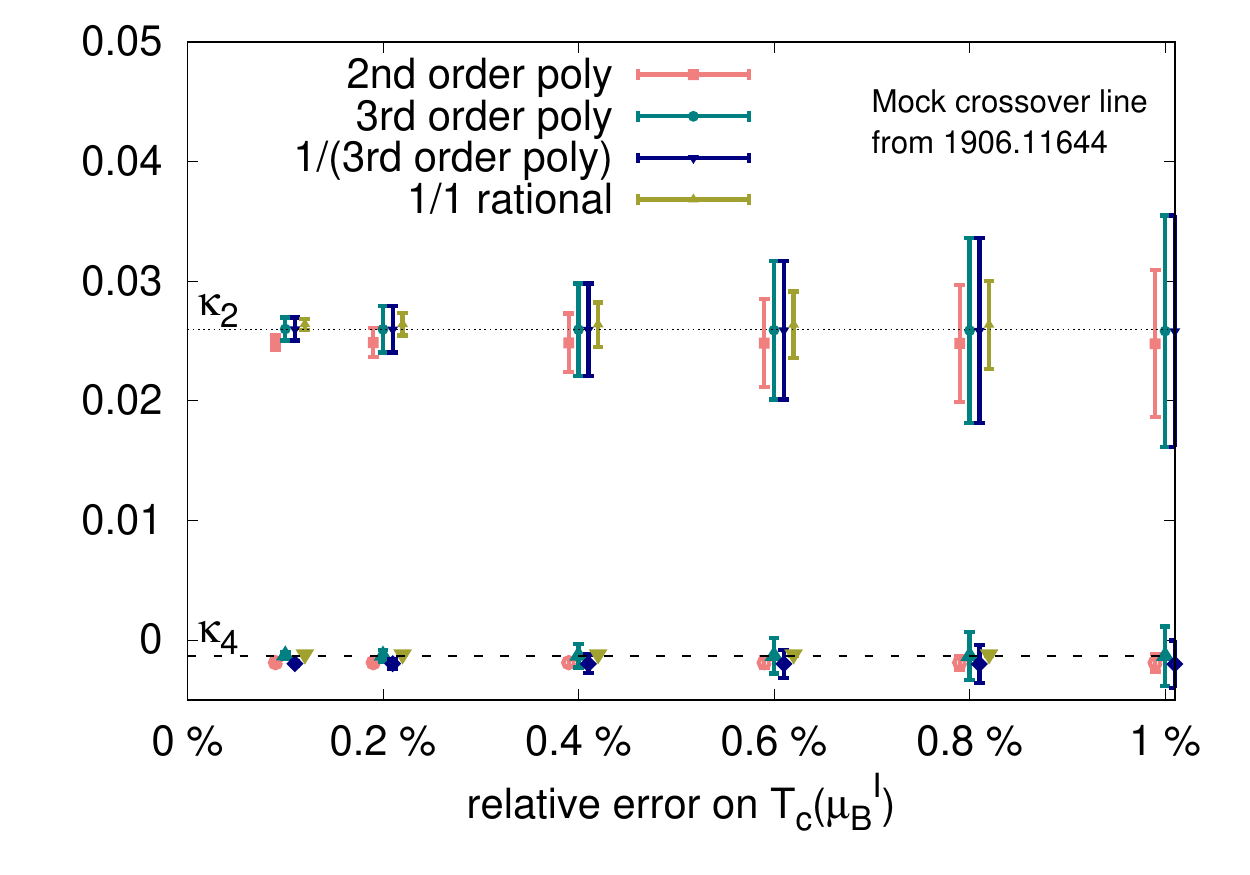}
\includegraphics[width=3.5in]{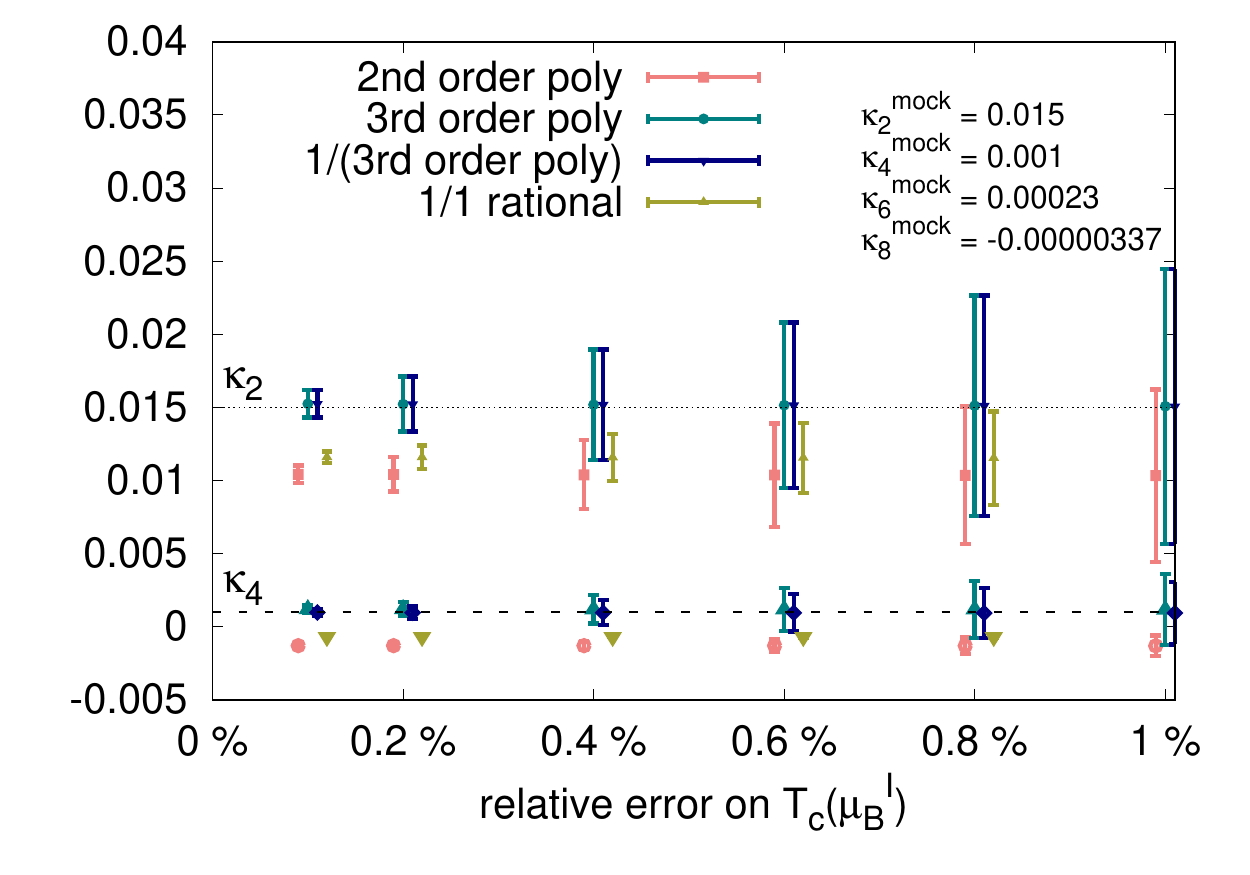}
\end{center}
\caption{\label{fig:mock}
Results for $\kappa_2$ and $\kappa_4$ obtained by fitting the mock data sets 
from model one (left panel) and model two (right panel). Each symbol corresponds to 
one of the fit functions described in the text. On the $x-$axis we show the relative
precision on the mock $T_c$. We marked the ``true'' $\kappa$ parameters of the mock
data with dashed lines. For small negative values of $\kappa_4$ all functions
perform well, for positive values only the third order functions give
correct coefficients.
}
\end{figure*} 

In the main text we calculated the transition temperature with
sub-percent precision. This precision is needed when we are attempting
to calculate a numerical derivative of $T_c$ with respect to
the chemical potential. The relevance of the sub-percent errors
in the $T_c$ determination can be highlighted in a mock analysis that
we present below.

We took two model scenarios for the crossover line and generated mock
data at imaginary chemical potential. Each mock data set 
was generated from a fourth order polynomial in $\hat{\mu}_B^2$
with coefficients $\kappa^{\rm mock}_2,\dots,\kappa^{\rm mock}_8$.  For the first model,
we fitted the cross-over line at real $\mu_B$ from Ref.~\cite{Isserstedt:2019pgx}, and
obtained the following values for the coefficients: $\kappa^{\rm mock}_2=0.0259463$, $\kappa^{\rm mock}_4=-0.0013438$, $\kappa^{\rm mock}_6=0.000053$ and $\kappa^{\rm
mock}_8=-0.00000094$. We then generated seven sets of mock data for each value of
imaginary chemical potential listed in Eq.~\eqref{eq:immu}, assigning to each set a fixed
relative error, ranging from $0.1\%$ to $1\%$ (see the left panel of Fig.~\ref{fig:mock}). 
We fitted each set of mock data $T_c(\hmu_B)/T_c(0)$ with four fit functions: a 2nd 
order polynomial, a 1/1 rational function $(1+ax)/(1+bx)$, a third order polynomial
and its reciprocal. $\kappa_2$ and $\kappa_4$ were extracted as the
leading Taylor coefficients of the fit functions. In the case of the two-parameter 
fits we dropped the largest imaginary chemical potential, so that all fits had four  
degrees of freedom. 

We repeated the whole procedure with a second model, constructed with 
a $\kappa^{\rm mock}_4$ parameter of opposite sign and same order of 
magnitude compared to the first one. For $\kappa^{\rm mock}_2$ and
$\kappa^{\rm mock}_4$ we used plausible values from the literature:
$\kappa_2^{\rm mock}=0.015$, $\kappa^{\rm mock}_4=0.001$, while we obtained 
$\kappa^{\rm mock}_6=0.00023$ and $\kappa^{\rm mock}_8=-0.00000337$ by forcing
the cross-over line to take the temperature values $205$~MeV, $157$~MeV and 
$100$~MeV at $\hmu_B=$ $i\pi$, 0 and 3, respectively
(see the right panel of Fig.~\ref{fig:mock}).

The purpose of the analysis was to determine whether the fits can reproduce 
the $\kappa_2$ and $\kappa_4$ values from the mock data. We found that the 
two models behave quite differently, even for large errors. In the first one, all 
the fit functions reproduce the value of $\kappa_4$.
In the second model, however, the second order fits yield $\kappa_4\approx -0.001$,
while its actual value is $\kappa^{\rm mock}_4=0.001$. The third order
fits perform much better. For the second model,
we find $\kappa_4 = 0.00123(18)$ and $\kappa_4 = 0.00099(16)$ from the per-mill 
precision mock data. However, with more precise data or a substantially higher 
$\kappa_8$ value, the third order fits would fail as well. In order for
$\kappa_8\hmu_B^8$ to be negligible relative to $\kappa_6\hmu_B^6$ in a fit
where $\hmu_B^2\in [-\pi^2,0]$, the coefficients $\kappa_8$ and $\kappa_6$
must be separated by more than an order of magnitude. 

Our real lattice data have a relative precision on $T_c(\mu_B)$ near
2 per-mills. Thus, in our analysis it is essential to take at
least third order polynomials to extract the cross-over line up to order $\hmu_B^4$.

Note also that, by reducing the relative error on $T_c(\hmu_B)$ from $0.5 \%$ to 
$0.2\%$, we significantly improved the 2nd order result on $\kappa_2$. 
However, when switching to 3rd order fit functions the error bars become as large 
as before. What was gained with the increased precision is an access to $\kappa_4$. 
This feature of deducing Taylor coefficients from polynomial fits is also pointed out 
in the main text.

\

\subsection{\label{sec:error}Error estimate}

In most results of this work, the statistical errors dominate. They were calculated
through the standard jackknife procedure with 48 bins. For each lattice size,
temperature and chemical potential we produced several (uncorrelated)
Monte Carlo streams. In most cases we calculated the chiral observables
after every five double-length trajectories. The sequence of measurements
in the Monte Carlo streams with equal bare parameters are concatenated
(after skipping 1000 trajectories for thermalization). 
The 48 bins are formed by splitting up this sequence evenly. 

Systematic errors are introduced every time we make an ambiguous choice in the
analysis. Such choices include the scale setting variable, which can be either
$f_\pi$ or $w_0$, the interpolating function we fit on the zero temperature
data and the one we use to fit the renormalized quantities. At several points
in the analysis we must make various decisions -- e.g. how many data points to
include into a fit or what the specific interpolating function should be. Each
time we pursue two or more choices, this results in splitting the analysis. The
final result is then calculated in many slightly different versions. The width
of their distribution yields the systematic error. 

In this paper we give combined errors. For this purpose, 
we consider the cumulative distribution function (CDF) for observable $x$
\begin{equation}
\mathrm{CDF}(x) = \sum_j w_j \frac12 \left(1-\mathrm{erf}\left[
(m_j-x)/\sqrt{2\sigma^2_j}
\right]\right)
\end{equation}
where we built a weighted sum of a Gaussian CDF corresponding
to analysis $j$ with mean $m_j$ and variance $\sigma_j$
and $\sum_j w_j=1$. We use uniform weights with
the aforementioned cut in the fit quality $Q$. The upper (lower) edge of 
the error bar is defined as the value 
of the observable corresponding to the 84\% (16\%) level of the
cumulative distribution function.

\begin{figure*}[ht]
\begin{center}
\hbox to \hsize{
\includegraphics[width=2.34in]{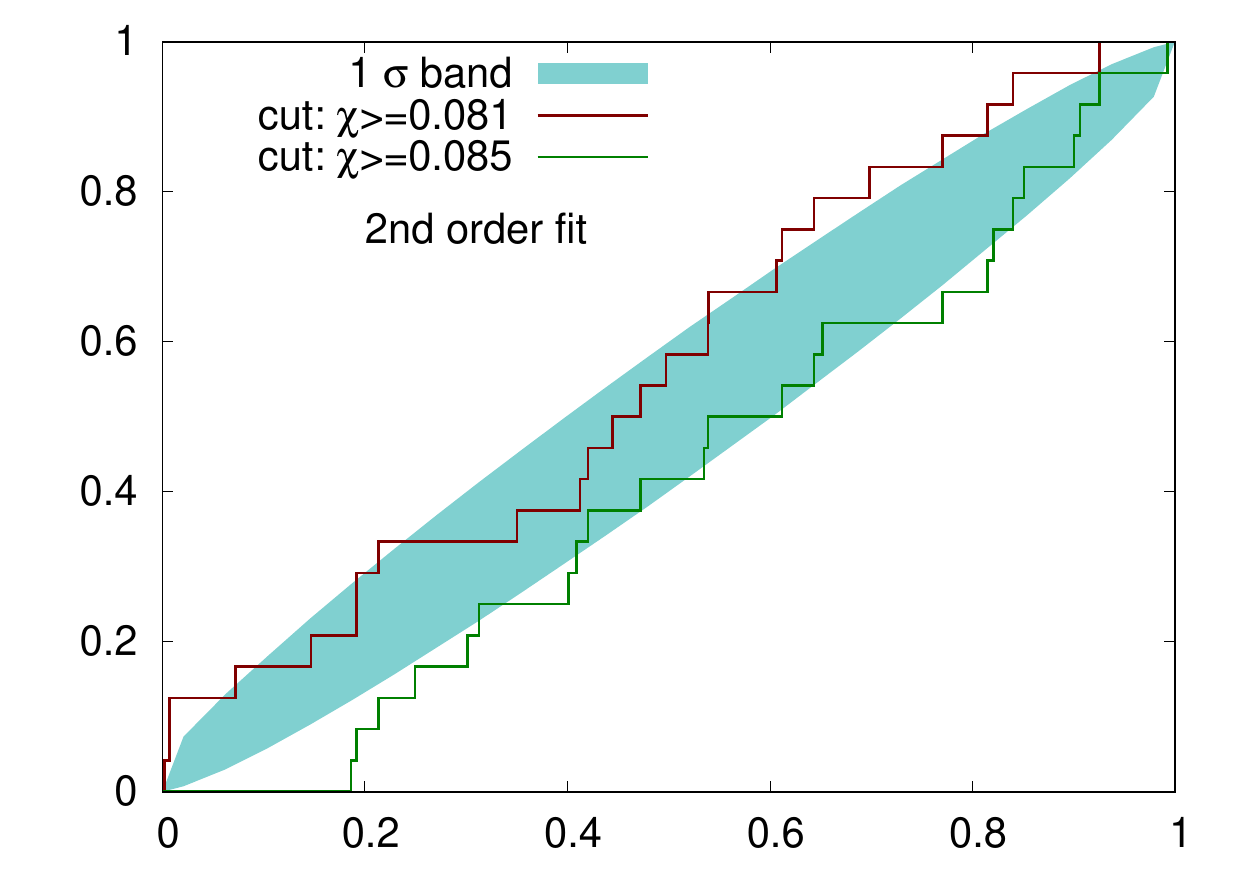}
\includegraphics[width=2.34in]{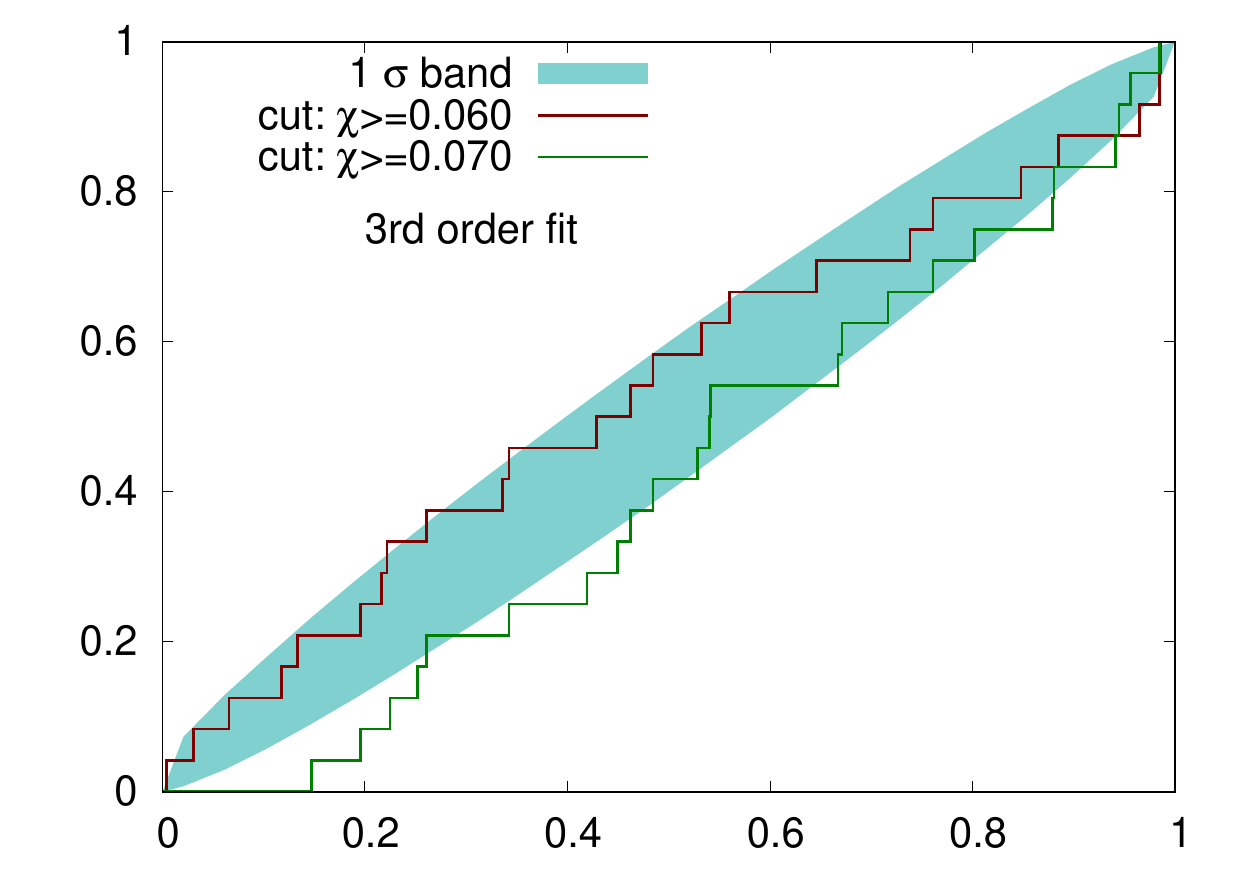}
\includegraphics[width=2.34in]{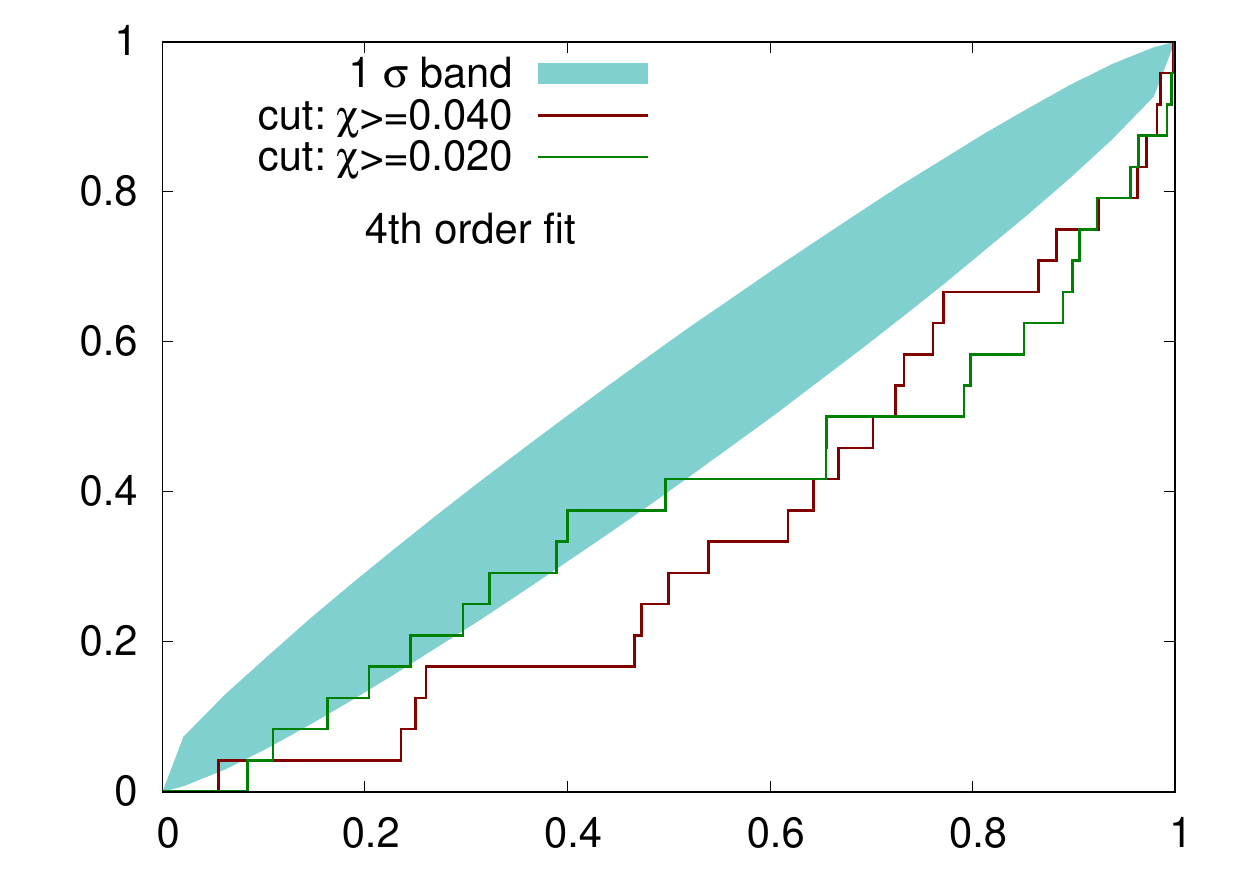}
\vspace{-0.8cm}
}
\end{center}
\caption{\label{fig:kstest}
Illustrations of the Kolmogorov-Smirnov test for our ${\chi}_{\rm cut}$ fit
range selection. For second (left), third (middle) and fourth (right) order
fits we selected two fit ranges for the determination of the peak position of
the chiral susceptibility. With the three lattice spacing and eight imaginary
chemical potentials we had 24 independent data sets for the test. The zig-zag
curves show the cumulative distribution function for of the $Q$ value of the
maximum fit, the almond shaped band shows the one-sigma expectation of the
curves. With higher $\chi_{\rm cut}$ less points are included, the polynomial
model might overfit the data, the corresponding curves go then below the almond.
Bad fits, on the other hand, are above the almond, this typically happens if
$\chi_{\rm cut}$ is too low, the fit is too inclusive. In the case of the
fourth order polynom always slightly overfit the data, thus we do not fit quartic polynomials.
}
\end{figure*}

\begin{figure}
\begin{center}
\includegraphics[width=3.5in]{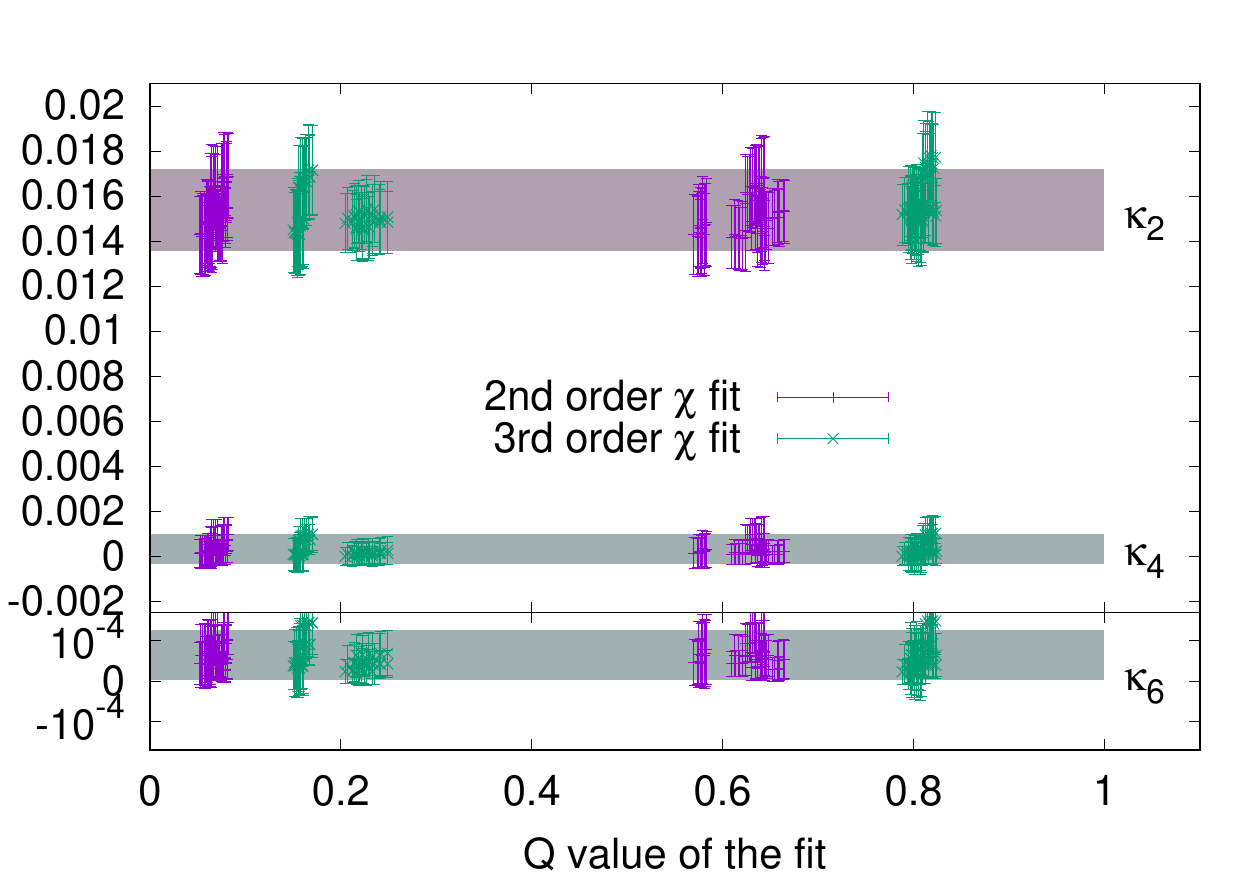}
\end{center}
\caption{\label{fig:kappaQ}
Results for $\kappa_2$ (upper set), $\kappa_4$ (middle set) and $\kappa_6$
(lower set) as functions of the fit quality. Each triplet of data points
corresponds to one analysis. The analyses are colored by the underlying 
($\chi$-peak fit order).
The final result (gray bands) is obtained by uniformly weighting all results.
}
\end{figure}

The different sources of systematic error in steps i-iv) of the analysis as described in the main text can be summarized as:
\setlist{nolistsep}
\begin{enumerate}[noitemsep, label=\roman*)]
    \item 2 choices of scale setting, 2 choices of the fit used for the renormalization of $\pbp$ and 2 choices of the fit used for the renormalization of $\chi$ as discussed in 
    the first supplement~\ref{sec:lat}

    \item The function $\chi(\pbp)$ is fitted with a quadratic or cubic polynomial on the range given by $\chi>\chi_{\rm cut}$. We choose two $\chi_{\rm cut}$
values for each polynomial order: we picked the largest
and smallest $\chi_{\rm cut}$ values that are endorsed by a Kolmogorov-Smirnov
test (see Fig.~\ref{fig:kstest}). 

    \item Two different cubic spline interpolations of $\pbp(T)$. The spline node points are selected to be at every other simulation temperature (the separation of our simulation temperatures is 5~MeV or less, see Fig.~\ref{fig:chinorm} (left) in the main text).
          Here we use either every even or every odd simulation point as node points.
    \item The global fit ansatz is either $T_c(\hmu_B^2,1/N_t^2)=\mathcal{F}(\hmu_B^2,1/N_t^2)$ or $T_c(\hmu_B^2,1/N_t^2)=\frac{1}{\mathcal{F}(\hmu_B^2,1/N_t^2)}$ with
          $\mathcal{F}(\hmu_B^2,1/N_t^2) = 1 + \hmu_B^2 ( a+ d/N_t^2) + \hmu_B^4 ( b+ e/N_t^2) + \hmu_B^6 ( c+ f/N_t^2)$ 
          with $\kappa_2=-a$, $\kappa_4=-b$ for the first fit and $\kappa_2=a$, $\kappa_4=b-a^2$ for the second. As a further source of systematic error, we either include the largest imaginary chemical potential in this fit, or we drop it.
\end{enumerate}
All these choices lead to the $2 \times 2 \times 2 \times 2 \times 2 \times 2 \times 2 \times 2 = 256$ 
analyses mentioned in the main text. The $Q$ values and the obtained $\kappa_2$ and $\kappa_4$ 
values for each fit are shown in Fig.~\ref{fig:kappaQ}. 
The fits are well spread in the $Q=[0,1]$ interval, and there is no
systematic dependence of the coefficients on the order of the used polynomial.
There are no too bad or too good fits, either, these were sorted out by
the Kolmogorov-Smirnoff tests of Fig.~\ref{fig:kstest}. 
This motivates us to combine the various analyses with a uniform weight.
The error is mainly statistical.  Among the statistical
effects the precision of the peak position in $\chi(\pbp)$ is the
bottleneck in this computation.

Notice our result on the $\kappa_6$ coefficient in the bottom of
Fig.~\ref{fig:kappaQ}. We cannot control the systematics of $\kappa_6$,
nevertheless, we find it takes a small value, consistent with
zero, and is stable for systematic effects. The order of magnitude
of $\kappa_6$ means that its effect on $T_c$ is not significant below 
$\mu_B\approx 300~\mathrm{MeV}$.

\subsection{\label{sec:strangeness} Strangeness neutrality} 

Throughout this work we use a strangeness chemical potential $\mu_S$,
which is always tuned such that for each simulated $\mu_B,\mu_S$ pair
we have strangeness density $n_S=0$. This is the strangeness
neutrality condition. We have already calculated the $\mu_S(\mu_B)$
dependence in Ref.~\cite{Bellwied:2015rza}, where we presented our
first continuum extrapolated result on the curvature $\kappa_2$ of
the transition line. 

As an improvement on our analysis from Ref.~\cite{Bellwied:2015rza}, we correct
the renormalized chiral condensate and susceptibility for the systematic or 
statistical deviations from the strangeness neutrality condition. To this end,
we calculate the derivatives of the condensate and the susceptibility with 
respect to the strangeness chemical potential. 

Let us first write the observables for $N_f$ flavors at fixed chemical potential:
\begin{eqnarray}
\pbp&=&\frac{T}{V} \langle b \rangle \, \, , \\
\chi&=&\frac{T}{V}\left[\langle a \rangle - \langle b\rangle^2\right] \, \, , \\
n_s&=&\frac{T}{V} \langle N_s\rangle \, \, ,
\end{eqnarray}
where:
\begin{eqnarray}
a&=&
m_l^2 N_f^2 (\mathrm{Tr}\, H^{-1})^2\nonumber\\
&&
+ N_f \mathrm{Tr}\, H^{-1}
-8m^2 N_f  \mathrm{Tr}\, \left( H^{-1} H^{-1} \right) \, \, ,\\
b&=& N_f m_l \mathrm{Tr}\, H^{-1} \, \, ,\\
N_s&=&\frac{d}{d\mu_s} \log (\det M_s)^{1/4} = \frac14\mathrm{Tr}\, \left( M_s^{-1} \right)
\frac{\partial{M_s}}{\partial {\mu_s}}\, .
\label{eq:Ns}
\end{eqnarray}
Here $H$ stands for the Hermitian matrix $H=M_l^\dagger M_l$ where $M_l$ is
the light quark matrix containing the light quark mass $m_l$ and
the light quark chemical potential $\mu_l =\frac13 \mu_B$. 
At the same time we introduce the strange
quark matrix $M_s$ containing the strange quark mass $m_s$ and the strange
quark chemical potential $\mu_s = \frac13 \mu_B -\mu_S$. The lattice observable $n_s$ is the strange quark density, $N_f = 2$ is the number of light flavors.
The factor 1/4~in front of the trace in Eq.~\eqref{eq:Ns} is due to the use of staggered quarks.

We then calculate:
\begin{eqnarray}
\frac{\partial\pbp}{\partial\mu_s} &=&
\frac{T}{V}\left[
\langle b \cdot N_s\rangle
-\langle b \rangle \langle N_s\rangle
\right]\,,\\
\frac{\partial\chi}{\partial\mu_s} &=&
\frac{T}{V}\left[
\langle a N_s\rangle - \langle a\rangle \langle N_s\rangle\right.\nonumber\\
&&\left.
- 2 \langle b\rangle \left(
\langle b N_s\rangle - \langle b\rangle \langle N_s\rangle
\right)
\right]\,,\\
\frac{\partial}{\partial \mu_j} \avr{N_s}&=& \avr{N_s^2}-\avr{N_s}\avr{N_s}+
\avr{ \frac{\partial N_s}{\partial \mu_s}}\,.
\end{eqnarray}
with
\begin{eqnarray}
\frac{\partial N_s}{\partial\mu_s}&=&  
\frac14\mathrm{Tr}
\,
\left(
\frac{\partial^2 M_s}{(\partial \mu_s)^2}  M_s^{-1}
-
\frac{\partial M_s}{\partial \mu_s}
M_s^{-1}
\frac{\partial M_s}{\partial \mu_s}
M_s^{-1}
\right)\,. ~~~
\label{eq:Nsmus}
\end{eqnarray}

We can correct for the small deviations from $n_s = 0 $ that we find
after performing the simulations, by calculating the change $\Delta\mu_s$ which would restore strangeness neutrality to leading order in a Taylor expansion on a jackknife-by-jackknife basis. We get the Taylor coefficients by averaging Eqs.~(\ref{eq:Ns}) 
and (\ref{eq:Nsmus}) ensemble-by-ensemble:
\begin{equation}
n_s + \Delta\mu_s \frac{\partial n_s}{\partial\mu_s} = 0\,,
~~\mathrm{solved~for}~\Delta\mu_s \, \, .
\end{equation}

We calculate the correction to the condensate and susceptibility to leading order in $\Delta\mu_s$
\begin{eqnarray}
\pbp_{\rm corr}= \pbp + \Delta\mu_s \frac{\partial\pbp}{\partial\mu_s}
\left(-\frac{m_l}{f_\pi^4}\right)\,\\
\chi_{\rm corr}= \chi + \Delta\mu_s \frac{\partial\chi}{\partial\mu_s}
\left(\frac{m_l^2}{f_\pi^4}\right)\,,
\end{eqnarray}
where $\pbp$ and $\chi$ are the renormalized observables from Eq.~\eqref{eq:chiobs} 
of the main text. Thus, for our corrected ensembles
strangeness neutrality is achieved with zero statistical error. The
$n_S=0$ setup may have a systematic error, though, if the leading order
expansion in $\Delta \mu_s$ was not satisfactory. However,
the ensembles were already tuned to fulfill $|n_S/N_B|< 0.05$ even without
correction. The correction $\Delta\mu_S/\mu_S<0.1$ introduced here
resulted in a relative shift of $\sim 10^{-3}$ or less for both $\pbp$ and $\chi$.

The correction that we calculated in this elaborate analysis is found to be
smaller than our statistical error in all cases, often by an order of
magnitude. Nevertheless, we applied the correction for all lattice spacings
that we use in this paper.

\subsection{Discussion of the width parameter}
A natural definition of the width of the transition is given by the curvature at the peak of the susceptibility, i.e.
\begin{equation}
(\Delta T)^2 = -\chi(T_c) \left[ \frac{d^2\chi}{dT^2} \right]^{-1}_{T=T_c} \, .
\label{eq:deltaT}
\end{equation}
The second derivative of $\chi(T)$ is numerically difficult to obtain, because it suffers from the systematic errors in the fit of $\chi(T)$. However, we can more easily model $\chi(\pbp)$ in a larger range of $\langle \pbp \rangle$.  
Thus, we introduce a proxy $\delta T$ for the half width $\Delta T$
\begin{eqnarray}
\delta T &=& 
\pbp^{-1} \left(
\pbp(T_c) + \frac{\Delta \pbp}2
\right)\nonumber\\
&&
-
\pbp^{-1} \left(
\pbp(T_c) - \frac{\Delta \pbp}2
\right) \, ,
\label{eq:deltaTalt}
\end{eqnarray}
where $\pbp^{-1}$ stands for the inverse function of $\pbp(T)$. 
If we define $\Delta \pbp$ analogously to Eq.~(\ref{eq:deltaT}) as
\begin{equation}
(\Delta \pbp)^2 = -\chi(\pbp) \left[ \frac{d^2\chi}{d\pbp^2} \right]^{-1}_{\pbp=\pbp_c} \, ,
\end{equation}
then using the fact that $\chi$ has a maximum at $T_c$, i.e.
\begin{equation}
\frac{d^2\chi}{dT^2}
= \frac{d^2\chi}{d\pbp^2} 
\left(\frac{d\pbp}{dT}\right)^2 \, \, ,
\end{equation}
we find that $\delta T = \Delta T$ up to higher order corrections in $\Delta \pbp$.

The values for this proxy are shown in Table~\ref{tab:tcimmu}. The proxy $\sigma$ 
used in the main text -- given by Eq.~\eqref{eq:Wdef} --  is even simpler. There, we fixed $\Delta \pbp$ to its $\mu_B=0$ value for all chemical potentials. This is justified by the last column of Table~\ref{tab:tcimmu}: the value $\Delta \pbp = 0.14$ is consistent with all the values in the range. For this reason, at $\mu_B=0$ we have
\begin{equation}
 \Delta T \approx \delta T \approx \sigma \, \, .
\label{eq:dTdTW}
\end{equation}

\begin{table}[!t]
\begin{center}
\vspace{4mm}
\begin{tabular}{c||c|c|c|c|c}
$\mathrm{Im}~\hmu_B$&$T_c$~[MeV] & $\delta T$~[MeV] & $\chi(T_c)$ & $\pbp(T_c)$ & $\Delta \pbp$\\
\hline\hline
0.000 & 158.01(61) & 14.9(0.9) & 0.129(2) & 0.284(3) & 0.139(7) \\
0.785 & 159.53(58) & 14.4(1.0) & 0.125(3) & 0.290(6) & 0.138(7) \\
1.178 & 161.14(57) & 15.0(0.9) & 0.129(3) & 0.282(5) & 0.148(7) \\
1.571 & 163.57(34) & 13.5(0.6) & 0.130(3) & 0.283(5) & 0.136(9) \\
1.963 & 166.67(78) & 14.9(1.0) & 0.129(3) & 0.287(7) & 0.148(8) \\
2.356 & 171.64(88) & 16.2(1.3) & 0.124(3) & 0.289(8) & 0.155(12) \\
2.553 & 174.72(83) & 14.7(1.4) & 0.131(4) & 0.296(6) & 0.132(12) \\
2.749 & 177.84(84) & 12.5(1.4) & 0.132(4) & 0.295(8) & 0.132(10) \\
\hline
\end{tabular}
\end{center}
\caption{\label{tab:tcimmu}
Continuum extrapolated values of the cross-over temperature $T_c$, the half 
width of the susceptibility peak $\delta T$, the height of the susceptibility
peak $\chi(T_c)$, the chiral condensate $\pbp (T_c)$ and the width parameter
$\Delta \pbp$ from our $\mathrm{Im}~\hmu_B$-by-$\mathrm{Im}~\hmu_B$ analysis.  Our final $\kappa_2$ and
$\kappa_4$ were obtained from a separate, 
more precise analysis, with correlated $T_c$ values.
We give this table to enable other researchers to use our data
for further model building.
}
\end{table}

\subsection{\label{sec:immudata}Tabulated continuum extrapolated results}

The extrapolation of $T_c(\mu_B)$ presented in this letter is the result of a 
global correlated fit. We could have pursued a different strategy: calculate 
the continuum extrapolated $\chi(\pbp)$ curves for each imaginary $\hmu_B$,
determine $T_c$ for each lattice at the given $\hmu_B$ and then
continuum extrapolate $T_c$ for fixed $\hmu_B$. This method has
the advantage to yield statistically uncorrelated $T_c(\hmu_B)$ values.
Its disadvantage is the lower precision: the simple polynomial
continuum extrapolation of the entire $\chi$ function has a better fit
quality if we restrict the data to a narrow peak region. Thus the best fits
have fewer data points. Nevertheless, we tabulate statistically independent
transition temperatures in Table~\ref{tab:tcimmu}, where the systematic
errors are already included. In this table we give the other
intermediate results as well: the width of the transition
$\delta T$, the peak height $\chi(T_c)$, the renormalized chiral condensate
$\pbp(T_c)$ and the width parameter $\Delta \pbp$.

In Table~\ref{tab:tcremu} we provide the  extrapolated cross-over temperature 
$T_c$ and the width parameter $\sigma$  at real $\mu_B$ from analytical continuation. They correspond to the green band in Fig.~\ref{fig:tcline} and the blue band in the upper panel of Fig.~\ref{fig:chi_immu}, respectively. Unlike in Table~\ref{tab:tcimmu}, the errors here are correlated, since these results are the output of a global analysis.

%

\begin{table}
\begin{center}
\vspace{4mm}
\begin{tabular}{c||c|c|c|c}
$\mu_B$~[MeV] & $T_c$~[MeV] & error~[MeV] & $\sigma$~[MeV] & error~[MeV] \\
\hline\hline
000 & 158.08 & 0.47 & 14.47 & 0.31 \\
020 & 158.03 & 0.48 & 14.46 & 0.32 \\
040 & 157.92 & 0.48 & 14.45 & 0.33 \\
060 & 157.72 & 0.48 & 14.43 & 0.36 \\
080 & 157.45 & 0.49 & 14.41 & 0.42 \\
100 & 157.08 & 0.51 & 14.36 & 0.50 \\
120 & 156.63 & 0.54 & 14.31 & 0.62 \\
140 & 156.08 & 0.60 & 14.23 & 0.79 \\
160 & 155.43 & 0.69 & 14.14 & 1.03 \\
180 & 154.67 & 0.83 & 14.01 & 1.36 \\
200 & 153.77 & 1.04 & 13.84 & 1.80 \\
220 & 152.72 & 1.35 & 13.61 & 2.41 \\
240 & 151.50 & 1.78 & 13.30 & 3.23 \\
260 & 150.04 & 2.39 & 12.89 & 4.35 \\
280 & 148.29 & 3.28 & 12.32 & 5.92 \\
300 & 146.16 & 4.62 & 11.49 & 8.22 \\
\hline
\end{tabular}
\end{center}
\caption{\label{tab:tcremu}
Continuum extrapolated cross-over temperature $T_c$ and the width parameter 
$\sigma$  at real $\mu_B$ from analytical continuation. The errors are the 
combined statistical and systematic errors discussed in the main text and 
the supplemental material.
}
\end{table}


\FloatBarrier


\newpage 

\bibliography{thermo}
\bibliographystyle{apsrev4-1}

\end{document}